\documentclass[reprint,aps,prd,superscriptaddress]{revtex4-2}

\usepackage{graphicx}
\usepackage{comment}
\usepackage{hyperref}
\usepackage{amsmath,amssymb,amsfonts}
\usepackage{graphicx}
\usepackage{dsfont}
\usepackage{ulem}
\usepackage{color}
\usepackage{physics}
\usepackage{bbold}
\usepackage{nccmath}
\usepackage{xcolor}

\usepackage{orcidlink}
\hypersetup{colorlinks=true}

\begin{document}


\title{Ultrasound response to time-reversal symmetry breaking below the superconducting phase transition}
\author{Chris Halcrow\orcidlink{0000-0002-0246-8075}}
\affiliation{Centre for Brain Discovery Sciences, Hugh Robson Building, University of Edinburgh, EH8 9XD, United Kingdom}
\affiliation{Department of Physics, KTH Royal Institute of Technology, 10691 Stockholm, Sweden}
\author{Paul Leask\orcidlink{0000-0002-6012-0034}}\email{palea@kth.se}
\affiliation{Department of Physics, KTH Royal Institute of Technology, 10691 Stockholm, Sweden}
\author{Egor Babaev\orcidlink{0000-0001-7593-4543}}\email{babaev@kth.se}
\affiliation{Department of Physics, KTH Royal Institute of Technology, 10691 Stockholm, Sweden}
\affiliation{Wallenberg Initiative Materials Science for Sustainability, Department of Physics, KTH Royal Institute of Technology, 10691 Stockholm, Sweden}
\date{\today}

\begin{abstract}
Ultrasound attenuation is a powerful probe of symmetry-breaking phenomena in superconductors.
In this work, we develop a framework to model the ultrasound response of multi-component superconductors undergoing a time-reversal symmetry breaking transition below the superconducting phase transition.
By coupling the elastic strain of the crystal lattice to the superconducting order parameters through group-theoretical analysis of tetragonal crystals, we classify how different symmetry channels contribute to the ultrasound signal.
Using a two-component Ginzburg--Landau theory, we analyze the temperature dependence of sound velocity   across both superconducting and time-reversal symmetry breaking transitions for several cases, including $(A_{1g}, A_{1g})$, $(A_{2g}, B_{1g})$, and $E_g$ representations.
Our results demonstrate that ultrasound measurements are highly sensitive to the presence of bilinear Josephson couplings and can distinguish between different realizations of the superconducting state.
We further show how external strain can significantly alter the ultrasound response in systems breaking time reversal symmetry.
\end{abstract}

\maketitle


\section{Introduction}

Ultrasound has long served as a powerful experimental tool for probing the properties of superconducting materials.
A peak in ultrasonic attenuation below $T_C$ was predicted by BCS (Bardeen--Cooper--Schrieffer) theory \cite{bardeen1957theory} and was confirmed  experimentally \cite{Tinkham_2004}.
In recent years, the scope of superconductivity research has expanded to especially focus on materials that exhibit multiple broken symmetries, particularly those with multi-component order parameters, including many unconventional pairing states  (see, e.g. \cite{Sigrist_1991}).
Identifying and understanding extra broken symmetries in a superconducting states is currently one of the central research theme in superconductivity.
Ultrasound is sensitive to the symmetry of the underlying superconducting state because it couples directly to lattice strain, which in turn interacts with the order parameter through symmetry-dependent couplings.
Variations in sound velocity and attenuation across phase transitions can thus reveal detailed information about the nature of symmetry breaking.
In particular, different components of the strain tensor transform according to distinct irreducible representations (irreps) of the crystal’s point group symmetry, allowing ultrasound to selectively probe specific symmetry channels.

Ultrasound data was one of the important probes that showed that UPt$_3$ \cite{muller1986observation} and UBe$_{13}$ \cite{golding1985observation} were unconventional superconductors, with interesting symmetries.
Ultrasound measurements have recently been used to argue that Sr$_2$RuO$_4$ is a multi-component superconductor \cite{benhabib2021ultrasound} and that URu$_2$Si$_2$ \cite{ghosh2020one} and UTe$_2$ \cite{Theuss_2024} have only one component.
In Ba$_{1-x}$K$_x$Fe$_2$As$_2$ the longitudinal and transverse ultrasonic responses  provided, along with other probes, evidence of a novel state: an electron quadrupling condensate \cite{grinenko2021state,Halcrow_2024}.
Furthermore, an unusual attenuation signal has been interpreted  evidence of domain walls, suggesting a broken additional discrete symmetry in Sr$_2$RuO$_4$ \cite{ghosh2022strong}.

One of the most interesting additional symmetries that a superconductor can break is a time-reversal symmetry (TRSB) \cite{Zhang2019,Li2011_2,Xia06,Vadimov2018,Maiti2013}.
Some superconductors undergo two transitions: from normal to time-reversal symmetric superconducting, and then to a TRSB phase.
Materials like Sr$_2$RuO$_4$, UPt$_3$, and Ba$_{1-x}$K$_x$Fe$_2$As$_2$ (at certain doping values $x$) were argued to exhibit such behavior.
Detecting and characterizing this transition is a central challenge in establishing new  kinds superconducting  states.

The main aim of this paper is to analyze and to give a detailed pedagogical description of how ultrasound attenuation and velocity change across superconducting phase transitions, particularly in materials that exhibit a lower-temperature TRSB transition within the superconducting phase.
We carry this out for materials that are described by mean-field theory.
To do this, we develop a theoretical framework using the  formalism similar to \cite{sigrist2002ehrenfest} and we model the coupled dynamics of strain and order parameters near the phase transitions.
We focus on tetragonal crystals with $D_{4h}$ symmetry, relevant to many candidate TRSB superconductors, and consider a variety of order parameter symmetries, including $(A_{1g}, A_{1g})$, $(A_{2g}, B_{1g})$, and $E_g$ irreps.
The ultrasonic wave's dispersion is modified by the interaction with the order parameter, leading to changes in sound velocity and attenuation (absorption coefficient).
We show that ultrasound can distinguish between these cases based on whether the sound velocity changes continuously or discontinuously at the transition, and whether responses appear in particular symmetry channels.

We discuss that ultrasound is not only sensitive to the presence of TRSB but also to further details of the structure of the superconducting order parameter, and the presence of bilinear Josephson couplings.
Since external stress can induce bilinear Josephson terms even in models where they are not present intrinsically,  combining that with ultrasound probes offers a tunable mechanism to explore symmetry-resolved properties of superconductors.

\begin{figure}[t]
    \centering
    \includegraphics[width=\linewidth]{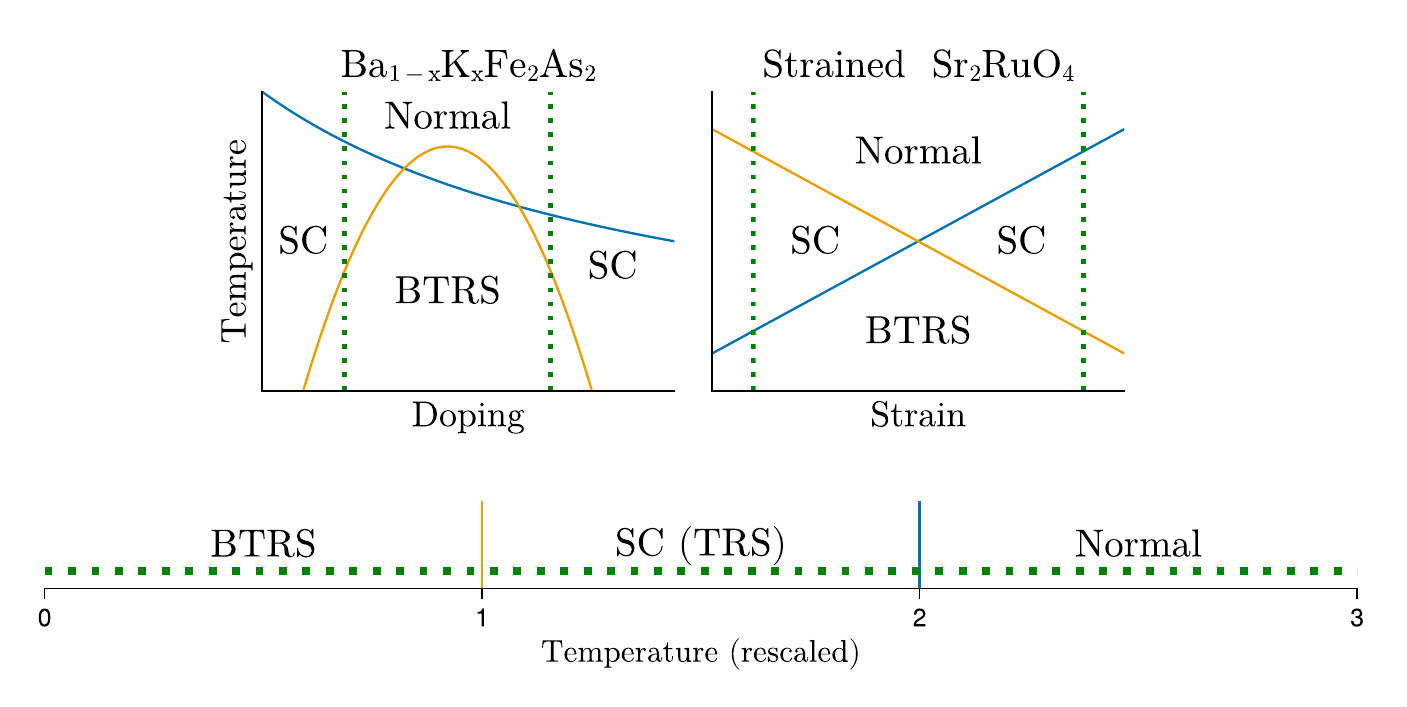}
    \caption{Schematic depictions of two discussed examples of phase diagrams of systems with a time-reversal symmetry breaking transition within the superconducting phase. The phase diagrams are schematic reconstructions for Ba$_{1-x}$K$_x$Fe$_2$As$_2$ (Fig. 1 in \cite{grinenko2021state}) and a discussed  phase diagram in  strained Sr$_2$RuO$_4$ (Fig. 1 in \cite{yuan2021strain}). All phase diagrams contain a 1D line where we transition from a normal to a time-reversal symmetric superconducting to a broken time-reversal phase as temperature decreases. This is the series of phase transitions we study in this paper.}
    \label{fig:phase_diagram}
\end{figure}


\section{Crystal symmetries and elastic strain}
\label{sec:strain}

To determine how ultrasound waves will respond to superconducting states in an elastic medium, we need to couple the superconducting theory to the strain of underlying crystal lattice,
\begin{equation}
\label{eq: Strain tensor}
    \epsilon_{ij} = \frac{1}{2} \left( \frac{\partial u_i}{\partial x_j} + \frac{\partial u_j}{\partial x_i} \right),
\end{equation}
where $\boldsymbol{u}$ is the local displacement vector.
We first begin by considering the ultrasound response to elastic deformations of the crystal lattice in the absence of superconductivity. 

We will consider a tetragonal crystal throughout this paper, but the analysis extends to cubic and orthorhombic crystals with small modifications.
Tetragonal crystals have a $D_{4h}$ symmetry group.
The order parameters and strain tensors can be decomposed into irreps of the group.
There are four one-dimensional irreps, $A_{1g}, A_{2g}, B_{1g}$ and $B_{2g}$, and one two-dimensional irrep $E_g$ \cite{Kaba_2019}.
The irreps, and some of the order parameters and strains that transform like them are displayed in Table \ref{tab: D4h irreps}.

\begin{table} [t]
    \begin{tabular}{l | c |  c | c | c }
		Irrep & Dim. & Order P.  & Strain & Elastic C. \\ \hline
		$A_{1g}$ & 1 & $s$, $d_{z^2}$  & $\epsilon_{xx} + \epsilon_{yy}$, $\epsilon_{zz}$ & $(C_{11}+C_{12})/2, C_{33}$ \\
		$A_{2g}$ & 1 & $g_{xy(x^2-y^2)}$  & --  & -- \\
		$B_{1g}$ & 1 & $d_{x^2-y^2}$ &  $\epsilon_{xx}-\epsilon_{yy}$ & $(C_{11}-C_{12})/2$ \\
		$B_{2g}$ & 1 & $d_{xy}$ & $\epsilon_{xy}$ & $C_{66}$ \\
		$E_g$ & 2  & $(d_{xz},d_{yz})$&  $(\epsilon_{xz}, \epsilon_{yz})$ & $C_{44}$\\
    \end{tabular}
    \caption{The irreducible representations of the $D_{4h}$ point group and their corresponding order parameters. Also shown in the right most two columns for each irrep is the strain tensor(s) that transform like that irrep, and their associated elastic moduli.}
\label{tab: D4h irreps}
\end{table}

Terms in the free energy must be invariant under all symmetries of the point group.
In representation theory language, all terms in the free energy must transform like $A_{1g}$.
To find the allowed terms which are quadratic in $\epsilon_{ij}$ we take the tensor product of each of the strain's irreps, with every other one.
The products of irreps can be calculated using a product table: the product table for $D_{4h}$ is shown in Tab. \ref{tab: Product Table}.
For example $\epsilon_{xy}$ transforms like $B_{2g}$.
The product table shows us that $B_{2g}\otimes B_{2g} = A_{1g}$.
The appearance of $A_{1g}$ means that there is one term made from two copies of $\epsilon_{xy}$ which appears in the free energy. 

In total there are six allowed terms, and the elastic free energy $\mathcal{F}_E$ can be written as
\begin{align}  \label{eq:Cirreps}
    \mathcal{F}_E = \, & \tfrac{1}{2}C_{A_{1g},1}\left( \epsilon_{xx} + \epsilon_{yy} \right)^2 + \tfrac{1}{2}C_{A_{1g},2}\epsilon_{zz}\left( \epsilon_{xx}+\epsilon_{yy}\right) \nonumber \\
    \, & + \tfrac{1}{2}C_{A_{1g},3}\epsilon_{zz}^2 
    + \tfrac{1}{2}C_{B_{1g}}(\epsilon_{xx} - \epsilon_{yy})^2 + \tfrac{1}{2}C_{B_{2g}}\epsilon_{xy}^2 \nonumber \\
    \, & + \tfrac{1}{2}C_{E_g} \left( \epsilon_{xz}^2 + \epsilon_{xy}^2 \right).
\end{align}
Here, each elastic constant is labeled by the irrep which also labels the underlying strain components.
This form highlights the connection between the irreps of $D_{4h}$ and the elastic energy.
The energy is more commonly written in tensor notation
\begin{equation}
    \mathcal{F}_E = \frac{1}{2}\epsilon_{ij}C_{ijkl}\epsilon_{kl},
\end{equation}
whose form makes applying symmetry operations simple.
For tetragonal crystals (with $D_{4h}$ point group symmetries), the elastic 4-tensor $C_{ijkl}$ has six components.
It is also common to switch to Voigt notation, where the elastic 4-tensor is written in terms of a $6\times 6$ matrix $C_{ij}$ with
\begin{align*}
    &C_{11} = C_{1111}, \quad C_{12} = C_{1122}, \quad C_{13} = C_{1133} \\
    & C_{33} = C_{3333}, \quad C_{66} = C_{1212}, \quad C_{44} = C_{1313}  ,
\end{align*}
plus symmetric permutations \cite{Luthi05}.
For a $D_{4h}$ symmetric system, twice the strain energy in Voigt notation is 
\begin{align} \label{eq:Cvoigt}
    2\mathcal{F}_E = \, &\frac{C_{11}+C_{12}}{2}\left( \epsilon_{xx} + \epsilon_{yy}\right)^2 + C_{13}(\epsilon_{xx}+\epsilon_{yy})\epsilon_{zz} \nonumber \\ 
    \, & + C_{33}\epsilon_{zz}^2 +\frac{C_{11}-C_{12}}{2}\left( \epsilon_{xx} - \epsilon_{yy}\right)^2 + C_{66}\epsilon_{xy}^2 \nonumber \\
    \, & + C_{44}\left( \epsilon_{xz}^2 + \epsilon_{xy}^2 \right),
\end{align}
which matches the irrep form \eqref{eq:Cirreps}.
We see that, for example, $C_{66} = C_{B_{2g}}$.
Hence $C_{66}$ is often referred to as the $B_{2g}$ elastic component.

\begin{table}[t]
    \centering
    \begin{tabular}{c|ccccc}
        $\otimes$&  $A_{1g}$& $A_{2g}$ & $B_{1g}$ & $B_{2g}$ & $E_g $\\ \hline
        $A_{1g}$  & $A_{1g}$ & $A_{2g}$ & $B_{1g}$ & $B_{2g}$ & $E_g $ \\
        $A_{2g}$  &  & $A_{1g}$ & $B_{2g}$ & $B_{1g}$ & $E_g$ \\
        $B_{1g}$  &  &  & $A_{1g}$ & $A_{2g}$ & $E_g$ \\
        $B_{2g}$  &  &  &  & $A_{1g}$ & $E_g$ \\
        $E_g $ &  &  &  &  & $A_{1g} \oplus B_{1g} \oplus B_{2g}$  \\
    \end{tabular}
    \caption{The product table for the even parity irreps of $D_{4h}$. We have disregarded antisymmetric elements.}
    \label{tab: Product Table}
\end{table}

The ultrasound response in the absence of superconductivity is straightforward to determine.
The linear elastic Lagrangian associated to the local displacement vector $\boldsymbol{u}(\boldsymbol{x},t)$ is defined by
\begin{equation}
    L = \frac{1}{2} \int \textup{d}\boldsymbol{x}\left( \sum_{i}\rho \dot{u}_i^2 - \sum_{ijkl} C_{ijkl}\frac{\partial u_i}{\partial x_j}\frac{\partial u_k}{\partial x_l} \right).
\end{equation}
The free ultrasound equations are most easily written using the 4-tensor notation.
These are obtained from Hamilton's principle
\begin{equation}
    \delta \int \textup{d}t L = 0 \quad \Rightarrow \quad \rho \frac{\partial^2 u_i}{\partial t^2} = C_{ijkl}\frac{\partial^2 u_l}{\partial x_j \partial x_k},
\end{equation}
with solutions given by the ansatz
\begin{equation}
    \boldsymbol{u}(\boldsymbol{x}, t) = \hat{\boldsymbol{u}} e^{i(\boldsymbol{k}\cdot \boldsymbol{x} -\omega t)}.
\end{equation}
where $\hat{\boldsymbol{u}}$ is a constant vector. We can now ask: which combinations of $\hat{\boldsymbol{u}}$ and $\boldsymbol{k}$ can be used to probe the different components of the strain tensor?
Consider the transverse wave $\hat{\boldsymbol{u}} = (1,0,0)$ and $\boldsymbol{k} = (0,1,0)$.
In this case, $\boldsymbol{u}$ only depends on $y$ and so the ultrasound equations simplify significantly. The only independent equation is
\begin{align}
    &\rho\ddot{u}_1 - C_{12k2}u_{k,22} = \rho \ddot{u}_1 - C_{1212}u_{1,22} = 0\nonumber \\
    &\implies \rho \ddot{u}_1 - C_{66}u_{1,22} = 0.
\end{align}
So this $\hat{\boldsymbol{u}}, \boldsymbol{k}$ combination only probes $C_{66}$: the $B_{2g}$ elastic component.
Hence, for a tetragonal crystal the ultrasound wave with $\hat{\boldsymbol{u}} = (1,0,0)$ and $\boldsymbol{k} = (0,1,0)$ is labelled using the $B_{2g}$ irrep.
In this way, irreps connect strain combinations to elastic tensor coefficients to properties of waves.
Using this chain of connections, we can label wave properties using irreps.
Note that the connections depend on the symmetry of the material.
A cubic or orthorhombic crystal has different relations, and the same sound modes will have a different irrep classification.

As an example, the sound velocity $v_{B_{2g}}$ will refer to the sound velocity of the wave with $\hat{\boldsymbol{u}} = (1,0,0)$ and $\boldsymbol{k} = (0,1,0)$.
The other examples and the labels relating irreps, $\boldsymbol{k}, \hat{\boldsymbol{u}}$, and $C$ combinations are displayed in Table \ref{tab:ukuC}.
The map between concepts is not bijective: the $A_{1g}$ elastic combination $C_{11} + C_{12}$ cannot be probed by a single wave.
Relatedly, the commonly used longitudinal wave, with $\hat{\boldsymbol{u}} = (1,1,0)$ and $\boldsymbol{k} = (1,1,0)$, does not probe a single irrep.
Instead, it probes a combination of $A_{1g}$ and $B_{2g}$ irreps.
For these modes, we use labels that are not the names of irreps.
This is a quirk of the tetragonal symmetry of our system: in orthorhombic and cubic crystals there is a one-to-one correspondence between elastic components and sound modes. 

\begin{table} [t]
    \begin{tabular}{c | c |  c c | c }
        Label & C & $\boldsymbol{k}$ & $\hat{\boldsymbol{u}}$  & $u$ \\ \hline
        $B_{1g}$ & $C_{11}-C_{12}$ & $(1,1,0)$ & $(1,-1,0)$  & $\epsilon_{xx} - \epsilon_{yy}$ \\
        $B_{2g}$ & $C_{66}$ & $(1,0,0)$ & $ (0,1,0)$  & $\epsilon_{xy}$ \\
        $A_{1g}$ & $C_{33}$ & $(0,0,1)$ & $(0,0,1)$  & $\epsilon_{zz}$  \\ 
        $E_g$ & $C_{44}$ & $(1,1,0)$ & $ (0,0,1)$  & $\epsilon_{xz} + \epsilon_{yz}$  \\ \hline 
        ``L"& $C_{11}+C_{12}+2C_{66}$& $ (1,1,0)$ & $ (1,1,0)$ & -- \\ 
        ``$A$" & $C_{11}$ & $(1,0,0)$ & $(1,0,0)$ & -- \\
        $A_{1g}$ & $C_{11} + C_{12}$ & -- & -- & $\epsilon_{xx}+\epsilon_{yy}$
    \end{tabular}
    \caption{The link between elastic moduli, strains, polarization vectors, and wave-directions. Note that the ``$L$" (Longitudal) and ``$A$" sound modes do not isolate a single irrep, while one of the $A_{1g}$ irreps has no corresponding sound mode.} \label{tab:ukuC}
\end{table}


\section{Landau--Khalatnikov-type formulation of the ultrasound response}

In this work we study the ultrasound response at the level of the phenomenological Landau--Khalatnikov framework that relies on being a vicinity of a phase transition \cite{Varma_1986}.
The superconducting order parameter is modeled using mean-field Ginzburg--Landau theory and is coupled linearly to the strain $\epsilon_{ij}$ of the underlying crystal lattice.
We note that this approach does not apply for strongly fluctuating systems, where, in superconductors with time-reversal symmetry breaking the sequence of the phase transition can be altered, requiring a different formalism developed in \cite{Halcrow_2024}.
Also, fluctuations make phase transition even in a relatively weakly coupled $U(1)\times Z_2$ superconductor, being weakly first order \cite{Haugen2021first}. We do not take into account these effects here considering a weak-coupled superconductor that is well-described by a mean-field theory.

Consider a two-component superconducting model with order parameter $\boldsymbol{\Psi}=(\Psi_1,\Psi_2)$, and a free energy with linear strain coupling:
\begin{equation}
\label{eq: General free energy}
    \mathcal{F} = V(\boldsymbol{\Psi}) + \frac{1}{2}\epsilon_{ij}C_{ijkl}\epsilon_{kl} + \Gamma_{ij}(\boldsymbol{\Psi})\epsilon_{ij},
\end{equation}
where $V(\boldsymbol{\Psi})$ is the potential energy of the order parameters and $\Gamma(\boldsymbol{\Psi})$ contains the order parameter terms that couple to the strain.
The ground state configurations $(\boldsymbol{\Psi}^0,\boldsymbol{u}^0)$ are solutions of the static field equations
\begin{subequations}
\label{eq:EoM}
    \begin{align}
        \left.\frac{\partial}{\partial \Psi_a}\right|_{\boldsymbol{\Psi}^0}\left( V - \frac{1}{2}\Gamma_{ij}C_{ijkl}^{-1}\Gamma_{kl} \right)  = 0, \\
        \epsilon^0_{ij} = C^{-1}_{ijkl}\Gamma_{kl}(\boldsymbol{\Psi}^0).
    \end{align}
\end{subequations}
Naively, the four-tensor $C_{ijkl}$ does not have a unique inverse.
However, the symmetries $i \leftrightarrow j$ and $k \leftrightarrow l$ imply that it does, provided the inverse also satisfies the symmetries.

To study ultrasound, we first perturb around the ground state solution $(\boldsymbol{\Psi}^0,\boldsymbol{u}^0)$.
When the order parameters are non-zero, it is helpful to choose a gauge.
In our models, it will always be possible to take $\Im(\Psi_1)=0$.
We must then factor out perturbations which are gauge transforms.
One way to do this is to choose new coordinates $\{ \psi_a \}$, which are all orthogonal to the gauge transformation.
We can calculate this at any given ground state by calculating the Hessian at that point, finding the kernel (which will be the local gauge transformation), and calculating the orthogonal complement to the kernel.
The vectors in the complement are the physical degrees of freedom.
Having chosen the $\psi_a$, we can express the perturbations as $\Psi_a = \Psi^0_a + \psi_a$ and $u_i = u_i^0 + u_i^\textup{wv}$, where $\textup{wv}$ stands for ``wave". 

For the free energy \eqref{eq: General free energy}, expanding about the ground state $(\boldsymbol{\Psi}^0,\boldsymbol{u}^0)$ up to second order in the perturbations $(\boldsymbol{\psi},\boldsymbol{u}^\textup{wv})$ gives
\begin{equation*}
    \mathcal{F} \simeq \mathcal{F}^{(0)} + \mathcal{F}^{(1)} + \mathcal{F}^{(2)},
\end{equation*}
where
\begin{align*}
    \mathcal{F}^{(0)} = \, & V(\boldsymbol{\Psi}^0) + \frac{1}{2}\epsilon_{ij}^0 C_{ijkl}\epsilon_{kl}^0 + \Gamma_{ij}(\boldsymbol{\Psi}^0) \epsilon_{ij}^0, \\
    \mathcal{F}^{(1)} = \, & \frac{\partial}{\partial \Psi_a} \left.\left( V(\boldsymbol{\Psi}) + \epsilon_{ij}^0 \Gamma_{ij}(\boldsymbol{\Psi})\right)\right|_{\boldsymbol{\Psi}^0}\psi_a \nonumber \\
    \, & +\frac{1}{2} C_{ijkl} \left( \epsilon_{ij}^0 \epsilon_{kl}^\textup{wv} + \epsilon_{ij}^\textup{wv} \epsilon_{kl}^0 \right) + \Gamma_{ij}(\boldsymbol{\Psi}^0)\epsilon_{ij}^\textup{wv}, \\
    \mathcal{F}^{(2)} = \, & \frac{1}{2}\frac{\partial}{\partial \Psi_a}\frac{\partial}{\partial \Psi_b} \left. \left( V(\boldsymbol{\Psi}) + \Gamma_{ij}(\boldsymbol{\Psi}) \epsilon_{ij}^0 \right)\right|_{\boldsymbol{\Psi}^0}\psi_a\psi_b \nonumber \\
    \, & \frac{1}{2}C_{ijkl} \epsilon_{ij}^\textup{wv}\epsilon_{kl}^\textup{wv} + \epsilon_{kl}^\textup{wv} \frac{\partial}{\partial \Psi_a} \left.\Gamma_{ij}(\boldsymbol{\Psi})\right|_{\boldsymbol{\Psi}^0}\psi_a.
\end{align*}
From the field equations, it is clear that the first order term in the expansion is $\mathcal{F}^{(1)}=0$.
Denote the physical derivatives evaluated at the static solution as $f_{,a} = \frac{\partial f}{\partial \psi_a}\rvert_{\psi(\boldsymbol{\Psi}^0)}$.
Then the free energy to second order in the perturbations is
\begin{align}
    \mathcal{F}^{(2)} = \, & \frac{1}{2}  \left[V_{,ab}(\Psi^0) + \Gamma_{ij,ab} \partial_j u^0_i \right]\psi_a \psi_b + \frac{1}{2} C_{ijkl}\epsilon^\textup{wv}_{ij}\epsilon^\textup{wv}_{kl} \nonumber \\ 
    \, &  + \Gamma_{ij,a}(\Psi^0)  \epsilon_{ij}^\textup{wv} \psi_a.
\end{align}
The dynamics of both the perturbations of the displacement vector $\boldsymbol{u}^\textup{wv}$ and order parameter $\boldsymbol{\psi}$ are described by the equations of motion \cite{sigrist2002ehrenfest} 
\begin{equation}
\label{eq: Landau-Khalatnikov}
    \rho \frac{\partial^2 u^\textup{wv}_i}{\partial t^2} = \partial_j \left( \frac{\partial \mathcal{F}^{(2)}}{\partial(\partial_j u^\textup{wv}_i)} \right), \quad \frac{\partial \psi_a}{\partial t} = - \frac{1}{\tau_0} \frac{\partial \mathcal{F}^{(2)}}{\partial \psi_a}.
\end{equation}
Here, $\rho$ is the density of the mass.
The evolution of the order parameter is modeled using the Landau--Khalatnikov formulation of overdamped dynamics, assuming $\tau_0$ is a phenomenological relaxation time $\sim T_C$.
The ultrasound equations \eqref{eq: Landau-Khalatnikov} are then
\begin{subequations}
    \begin{align}
        \tau_0 \frac{\partial \psi_a}{\partial t} + \left[ V_{,ab} + \Gamma_{ij,ab}\partial_j u_i^0 \right]\psi_b + \Gamma_{ij,a}\epsilon_{ij}^\textup{wv} = \, & 0, \\
        \rho\ddot{u}^{\textup{wv}}_i - \partial_j \left( C_{ijkl}\epsilon_{kl}^\textup{wv} + \psi_a \Gamma_{ij,a} \right) = \, & 0.
\end{align}
\end{subequations}
These have plane wave solutions
\begin{equation}
    \psi_a = A_a e^{i(k_i x_i - \omega t)}, \quad u_i^{\textup{wv}} = \hat{u}_{i} e^{i(k_i x_i - \omega t)},
\end{equation}
with
\begin{align}
    A_a = \, & -i\hat{u}_i k_j \tilde{V}_{ab}^{-1}\Gamma_{ij,b}, \\
    \tilde{V}_{ab} = \, &  V_{,ab} + \Gamma_{ij,ab}\partial_j u_i^0 - i \omega \tau_0 \delta_{ab}.
\end{align}
This ansatz leads to a vector of dispersion relation
\begin{equation}
\label{eq:nonldisp}
     D_i \equiv \rho\omega^2 \hat{u}_{i} - \left(C_{ijkl}- \Gamma_{ij,a}\tilde{V}^{-1}_{ab} \Gamma_{kl,b} \right)k_j k_l \hat{u}_{k} = 0.
\end{equation}
The dispersion relation depends on the choice of initial wave through $\hat{\boldsymbol{u}}$ and $\boldsymbol{k} = k_{\mathcal{I}} \hat{\boldsymbol{k}}$.
Here, we have made the choice of wave explicit through the wave vector magnitude $k_{\mathcal{I}}$, where $\mathcal{I}$ is equal to any of the labels in Tab. \ref{tab:ukuC}.
Most pulse-echo experiments measure the sound velocity of the same wave that is put into the system.
We calculate this by projecting the (vector of) dispersion relations $\boldsymbol{D}$ onto the initial polarization $\hat{\boldsymbol{u}}$.
Then the following equation is a quadratic in $k_{\mathcal{I}}$:
\begin{equation}
\label{eq:nonldispone}
     D_i\hat{u}_{i}=\rho\omega^2 \hat{u}_{i}\hat{u}_{i} - (C_{ijkl} - \Gamma_{ij,a}\tilde{V}^{-1}_{ab} \Gamma_{kl,b})\hat{u}_{i}k_j\hat{u}_{k}k_l = 0,
\end{equation}
which can be solved for $k_I$ either numerically or analytically. The sound velocity and absorption coefficients are given by \cite{sigrist2002ehrenfest}
\begin{equation}
    v_\mathcal{I} = \frac{\omega}{\textup{Re } k_{\mathcal{I}}(\omega)},\quad \alpha_{\mathcal{I}} = -\textup{Im }k_{\mathcal{I}}(\omega).
\end{equation}
These are the quantities measured in pulse-echo experiments \cite{Williams_1970}.
In general, the background strain $\partial_j u_i^0$ and the relaxation time $\tau_0$ are small.
Hence the main contribution to ultrasound response comes from the Hessian $V_{,ab}$.
There are several possible sound modes to choose from.

The sound velocity in the normal phase is given by the same procedure but with the interaction term ($\Gamma V^{-1}\Gamma$) set to zero.
This defines the normal sound velocity for each label $v_{\mathcal{I}0}$.
It is equal to
\begin{equation}
    v_{\mathcal{I}0} = \sqrt{ C_{ijkl}\hat{u}_{i}\hat{k}_j\hat{u}_{k}\hat{k}_l} \equiv \sqrt{C_{\mathcal{I}0}} ,
\end{equation}
where we have defined the normal elastic constant for each label $\mathcal{I}$.
The $C_{\mathcal{I}0}$ are the ones listed in the $C$ column of Tab. \ref{tab:ukuC}.

In the large $C$, small $\tau_0$ limit the normalized change in sound velocity for an ultrasound wave with wavevector $\boldsymbol{k}$ and polarization $\boldsymbol{u}_0$ is given by
\begin{equation}
\label{eq:analytic_formula}
    \Delta \tilde{v}_\mathcal{I} =  C_{\mathcal{I}0}\frac{\Delta v_{\mathcal{I}}}{v_{\mathcal{I}0}} = -\frac{\hat{u}_{i}\hat{u}_{k} \hat{k}_j \hat{k}_l\Gamma_{ij,a}\tilde{V}^{-1}_{,ab}\Gamma_{kl,b}}{2}.
\end{equation}
The ultrasound response is only non-zero if the term $\Gamma_{ij,a}\tilde{V}^{-1}_{ab} \Gamma_{kl,a}\hat{k}_j \hat{k}_l \hat{u}_{k}$ is non-zero.
Again, this depends on the choice of the initial wave through $\boldsymbol{u}_0$ and $\boldsymbol{k}$ and the structure of $\Gamma_{ij}(\boldsymbol{\Psi})$, which in turn depends on the symmetries of the order parameter.
In resonant ultrasound spectroscopy (RUS) experiments, the change to the elastic tensor is measured \cite{ghosh2020one,Theuss_Simarro_2024}.
Its value is closely related to the change in sound velocity in the large $C$, small $\tau_0$ limit,
\begin{equation}
    \Delta C_{ijkl} = -\Gamma_{ij,a}\tilde{V}^{-1}_{,ab}\Gamma_{kl,b}.
\end{equation}


\section{Ultrasound response to broken time-reversal symmetry}

We now study a specific model in more detail.
Consider the free energy 
\begin{equation}
\label{eq:freeenergyFULL}
    \mathcal{F} = V(\boldsymbol{\Psi}) + \mathcal{F}_E(u) + \mathcal{F}_C(\boldsymbol{\Psi}, u),
\end{equation}
where the order parameter potential is quartic and is given by
\begin{align} \label{eq:VEnergy}
    V = \, & -a_1 |\Psi_1|^2 - a_2 |\Psi_2|^2  + \frac{b}{4}(|\Psi_1|^2 + |\Psi_2|^2)^2 \nonumber  \\
    \, & + c_1\left( \Psi_1 \Psi^{\dagger}_2 + \Psi^{\dagger}_1\Psi_2 \right)+ \frac{c_2}{2}\left( \Psi_1^2 \Psi^{\dagger2}_2 + \Psi^{\dagger2}_1\Psi_2^2 \right),
\end{align}
the tetragonal elastic energy $\mathcal{F}_E$ \eqref{eq:Cvoigt} discussed in Section \ref{sec:strain} and a term $\mathcal{F}_C$ which couples terms linear in $u$ and quadratic in the $\Psi_a$.
The $T$ dependence is approximated through $a_1$ and $a_2$,
\begin{equation}
    a_1 = -T+T_1, \quad a_2 = -T+T_2.
\end{equation}
Note that, due to nonzero $c_1$ and $c_2$, $T_1$ and $T_2$ are not the critical temperatures. 
The coupling term depends on the symmetries of the order parameters.
In the next three subsections, we consider three different models with different order parameter symmetries.
From now on we ignore strain in the $z$-direction, for simplicity.

In the absence of the bilinear and biquadratic Josephson terms, $\Psi_1 \Psi^{\dagger}_2 + \Psi^{\dagger}_1\Psi_2$ and $\Psi_1^2 \Psi^{\dagger2}_2 + \Psi^{\dagger2}_1\Psi_2^2$, the system has a $U(1) \times U(1)$ symmetry, since the phases $\theta_j$ of the two order parameters $\Psi_j=|\Psi_j|e^{i\theta_j}$ are independent in this case.
However, let us include the biquadratic Josephson term and express it as
\begin{equation}
    \frac{c_2}{2}(\Psi_1^2 \Psi^{\dagger2}_2 + \Psi^{\dagger2}_1\Psi_2^2) = c_2\lvert\Psi_1\rvert\lvert\Psi_2\rvert\cos[2(\theta_1-\theta_2)].
\end{equation}
If we take $c_2>0$, then the term $\cos[2(\theta_1-\theta_2)]$ is minimized for the choice $\theta_1-\theta_2=\pm\tfrac{\pi}{2}$.
This gives two possible ground states that are related by a complex conjugation operation, which introduces a $\mathbb{Z}_2$ symmetry into the model.
The choice of ground state breaks the $U(1)\times\mathbb{Z}_2$ symmetry and is known as time-reversal symmetry breaking.

We will focus on superconductors where the TRSB phase occurs below the superconducting phase, that is, we consider $T_C^{U(1)} > T_C^{\mathbb{Z}_2}$.
Above $T_C^{U(1)}$ we obviously have the normal phase.
The order parameter is different in each of the three distinct phases.
The bilinear Josephson term means it is not possible to find simple analytic expressions for the order parameters, though we can make progress near the critical points.

In the time-reversal symmetric $U(1)$ superconducting phase near the transition, both order parameters are real and small and proportional to $\sqrt{T-T_C^{U(1)}}$.
We use this approximation to determine the order parameters near $T_C^{U(1)}$, they are found to be given by
\begin{subequations}
\label{eq: OP U1}
    \begin{align}
        \Psi_1^2 &=\frac{2 a_1 c_1^2 \left(a_1 a_2-c_1^2\right)}{a_1^4 b+2 a_1^2 c_1^2 (b+2 c_2)+b c_1^4}, \\
        \Psi_2 &= \frac{a_1}{c_1}\Psi_1,
    \end{align}
\end{subequations}
and the potential energy near $T_C^{U(1)}$ reduces to
\begin{equation}
    V(T) = -\frac{a_2^2 \left( c_1^2 - a_1a_2 \right)^2}{b\left( a_2^2 + c_1^2 \right)^2 + 4a_2^2c_1^2c_2}.
\end{equation}
We can  express $T_C^{U(1)}$ in terms of $T_1$ and $T_2$, which yields
\begin{equation}
\label{eq:TC1}
    T_C^{U(1)} = \frac{1}{2}\left\{ T_1 + T_2 + \sqrt{4c_1^2 + (T_1-T_2)^2 } \right\}.
\end{equation}

In this model, the TRSB phase occurs when $a_2$ becomes large enough.
We can approximate this by taking the limit when the bilinear Josephson coupling is zero.
In this limit, the order parameter satisfies $\textup{Re}(\Psi_1\Psi^{\dagger}_2) = 0$.
Then the order parameters are given by
\begin{subequations}
\label{eq: OP TRSB}
    \begin{align}
        \Psi_1^2 = \frac{b(a_2-a_1) - 2a_2c_2}{2c_2^2 - 2bc_2 }, \\
        \Psi_2^2 = \frac{b(a_2-a_1) - 2a_1c_2}{2c_2^2 - 2bc_2 },
    \end{align}
\end{subequations}
and the TRSB critical temperature is found to be 
\begin{equation}
\label{eq:TC2}
    T_C^{\mathbb{Z}_2} = T_2 + \frac{b}{2c_2}(T_2-T_1).
\end{equation}

\begin{figure}[t]
    \centering
    \includegraphics[width=1.0\linewidth]{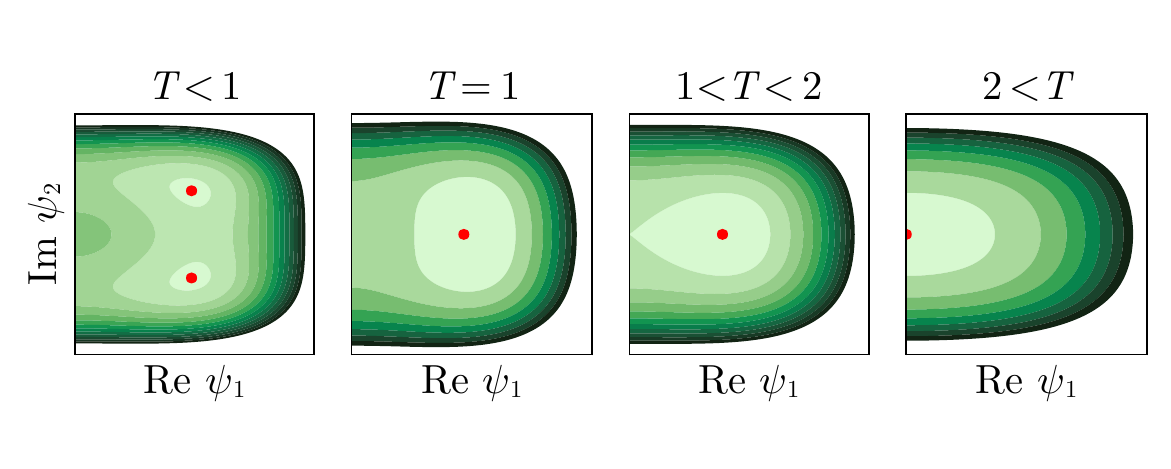}
    \caption{The ground state configuration of the order parameters $(\Psi_1,\Psi_2)$ in our model as a function of the temperature $T$. The temperatures $T_1$ and $T_2$ are normalized such that $T_C^{U(1)}=2$ and $T_C^{\mathbb{Z}_2}=1$. It is clear to see that there are two transitions: one from the normal phase ($T>2$) to the time-reversal symmetric $U(1)$ superconducting phase ($1<T<2$), and the second from the time-reversal symmetric to the broken time-reversal symmetry phase ($T<1$). It can be seen that below $T_C^{\mathbb{Z}_2}=1$ there are two possible ground states, which introduces a $\mathbb{Z}_2$ symmetry into the model, and breaks time-reversal symmetry.}
    \label{fig: Ground state}
\end{figure}

It will be convenient to normalize the critical temperatures, in terms of $T_1$ and $T_2$, such that $T_C^{U(1)}=2$ and $T_C^{\mathbb{Z}_2}=1$.
Using the relations for the critical temperatures, \eqref{eq:TC1} and \eqref{eq:TC2}, we can invert them to find the temperatures $T_1$ and $T_2$ such that the normalization of the critical temperatures is ensured.
They are found to be given by
\begin{subequations}
    \begin{align}
        T_1 = \, & \frac{-\sqrt{b^2 c_1^2+2 b c_1^2 c_2+c_2^2}+2 b+c_2}{b}, \\
        T_2 = \, & \frac{-\sqrt{b^2 c_1^2+2 b c_1^2 c_2+c_2^2}+2 b+3 c_2}{b+2 c_2},
    \end{align}
\end{subequations}
and we use these values for the rest of the paper. 
In what follows as a concrete example, we also use the potential parameters 
\begin{align*}
    (a_1,&a_2,b, c_2, \gamma_1, \gamma_2, \gamma_3) \\
    &= (-T+T_1, -T+T_2, 1, 0.3, 0.6, 0.8, 1.0) ,
\end{align*}
elastic parameters
\begin{align*}
    C_{11} = C_{33} = 100, \quad C_{12} = C_{13} = C_{44} = C_{66} = 10 
\end{align*}
and relaxation parameter $\tau_0 \omega = 0.001$.
The coefficient of the bilinear Josephson coefficient $c_1$ is either zero or non-zero depending on the symmetry of the order parameter.
The ground state configurations of the order parameter are shown in Fig.~\ref{fig: Ground state} in various temperature $T$ regimes.

\subsection{$(A_{1g}, A_{1g})$ order parameter}

Consider a two-component order parameter where each component transforms like $A_{1g}$.
A common example of this are superconductors with two $s$-wave components.
These are argued to describe iron-based superconductors, with evidence coming from angle-resolved photoemission spectroscopy (ARPES) experiments \cite{ding2008observation, nakayama2011universality,corbae2024fundamental}.
The bilinear Josephson term transforms as $A_{1g}$ and so it can appear in the free energy with $c_1\neq 0$; we take $c_1=0.2$.
All three terms in \eqref{eq:quadratics} transform as $A_{1g}\otimes A_{1g} = A_{1g}$.
Hence they all couple to the $A_{1g}$ strain, $\epsilon_{xx} + \epsilon_{yy}$, and the coupling term is
\begin{align}
\label{eq: (s,s) potential}
    \mathcal{F}_C^{(s,s)} = \, & \left[ \gamma_1\left( |\Psi_1|^2 + |\Psi_2|^2 \right) +  \gamma_2\left( |\Psi_1|^2 - |\Psi_2|^2 \right) \right.  \nonumber \\ 
    \, & \left.+ \gamma_3 \left( \Psi_1\Psi^{\dagger}_2 + \Psi^{\dagger}_1 \Psi_2 \right) \right] \left( \epsilon_{xx} + \epsilon_{yy} \right).
\end{align}

\begin{figure}
    \centering
    \includegraphics[width=\columnwidth]{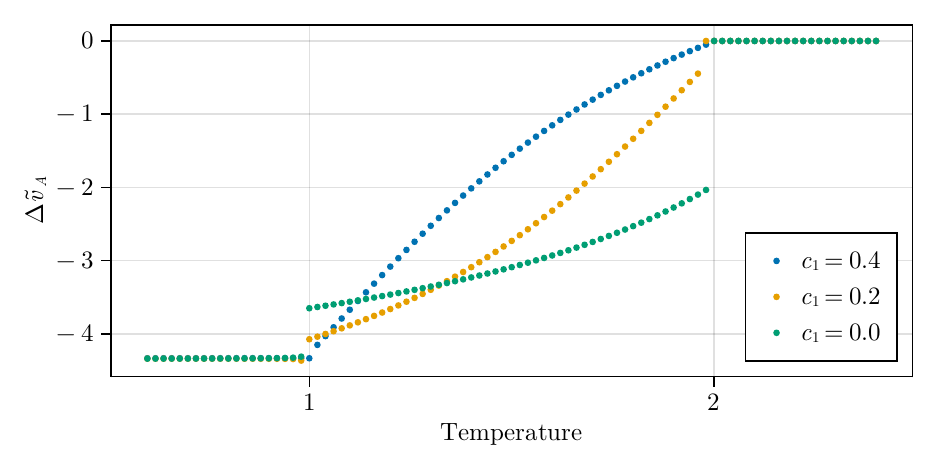}
    \caption{The $A$-mode response in the $(A_{1g}, A_{1g})$ model, corresponding to $(s,s)$-components, as a function of the bilinear Josephson coefficient $c_1$, showing that ultrasound is sensitive to the size of the bilinear Josephson term. Although the bilinear Josephson coupling alters the critical temperatures via the $c_1$ coefficient, we normalize them such that $T_C^{U(1)}=2$ and $T_C^{\mathbb{Z}_2}=1$. While the critical temperatures are normalized for comparison, the change in $c_1$ is absorbed into the temperatures $T_1$ and $T_2$. This allows us to compare the $A$-mode response as the coupling $c_1$ is increased. We observe that the discontinuities in the ultrasound response get smaller as $c_1$ increases. }
    \label{fig:biJ_dep}
\end{figure}

In Figure \ref{fig:biJ_dep} we plot the $A$-mode response for an $(A_{1g}, A_{1g})$ model for a variety of bilinear Josephson coefficients $c_1$.
We see that the ultrasound response is highly sensitive to the size of the coefficient of the bilinear Josephson term,  $c_1$.
The discontinuity in the responds becomes smaller as $c_1$ is increased.
Hence ultrasound can be used to constrain whether a theory does or doesn't have a bilinear Josephson term, and help to measure its size.

The ultrasound response for this model \eqref{eq: (s,s) potential} is shown in the left column of Figure \ref{fig:data_1}.
The only non-zero response occurs in the $A$ channel, where $\hat{\boldsymbol{u}} = \hat{\boldsymbol{k}} = (1,0,0)$.
The response in the superconducting phase ($1<T<2$) is almost linear, and there are only small jumps at the transitions to the normal phase ($T>2$) and to the TRSB phase ($T<1$).

\subsection{$(A_{2g}, B_{1g})$ order parameter}
\label{subsec: A2g times B1g}

Now consider a two-component superconductor where one of the order parameters transforms like $A_{2g}$ and the other like $B_{1g}$.
This choice is taken from the recent proposal of \cite{kivelson2020proposal} to describe Sr$_2$RuO$_4$ using a $g_{xy(x^2-y^2)}+id_{xy}$ pair of order parameters.
Since $A_{2g}\otimes A_{2g} = B_{1g} \otimes B_{1g} = A_{1g}$, the squares of the order parameters couple to the $A_{1g}$ strain, $\epsilon_{xx} + \epsilon_{yy}$.
However, using Table \ref{tab: Product Table}, the product $\Psi^{\dagger}_1 \Psi_2$ transforms as $A_{2g} \otimes B_{1g} = B_{2g}$.
Hence the term $\Psi^{\dagger}_1 \Psi_2 + \Psi_1\Psi^{\dagger}_2$ couples to the $B_{2g}$ strain, $\epsilon_{xy}$, since $B_{2g} \otimes B_{2g} = A_{1g}$.
The coupling term then takes the form
\begin{align}
   \mathcal{F}_C^{(g,d)} = &\left[ \gamma_1\left( |\Psi_1|^2 + |\Psi_2|^2 \right) + \gamma_2\left( |\Psi_1|^2 - |\Psi_2|^2 \right)  \right]  \\
    &\times \left( \epsilon_{xx} + \epsilon_{yy} \right) \nonumber + \gamma_3 \left( \Psi_1\Psi^{\dagger}_2 + \Psi^{\dagger}_1 \Psi_2 \right) \epsilon_{xy}. 
\end{align}
There is no bilinear Josephson since this term transforms as $B_{2g}$ and not $A_{1g}$; meaning that $c_1 = 0$. 

The ultrasound response for this model is shown in the middle column of Figure \ref{fig:data_1}. Two of the three sound modes have a non-zero response.
Most interestingly, the $B_{2g}$ response is continuous.
The change in sound velocity at the superconducting transition is given by
\begin{equation}
\label{eq: B2g continuous velocity change}
    \Delta \tilde{v}_{B_{2g}} =  \frac{\gamma_3^2 (-T+T_C^{U(1)})}{a_2b},
\end{equation}
which is linear in $T-T_C^{U(1)}$, explaining the continuity.
Although there is no bilinear Josephson term in the free energy, it is interesting to see what its effect would have been.
So we now insert a non-zero $c_1$ and re-evaluate the change in sound velocity. We can expand in small $T-T_C^{U(1)}$, $c_1$ and $c_2$ to find
\begin{equation}
\label{eq: B1g discontinuous change}
    \Delta \tilde{v}_{B_{1g}} = -\gamma_3^2\left( \frac{ 4c_2}{b^2} - \frac{1}{b} \right) ,
\end{equation}
which is a discontinuous jump in the response.
Hence the result is only continuous provided $c_1=0$.
This is the opposite of the effect of $c_1$ on the $(s,s)$-model.
There, $c_1$ was the cause of smoothness, here it causes discontinuity.
Again, the bilinear Josephson term is an essential term to consider in ultrasound calculations.
A similar result holds at the TRSB transition.


\subsection{$E_g$ order parameter}

\begin{figure*}[t]
    \centering
    \includegraphics[width=1.0\linewidth]{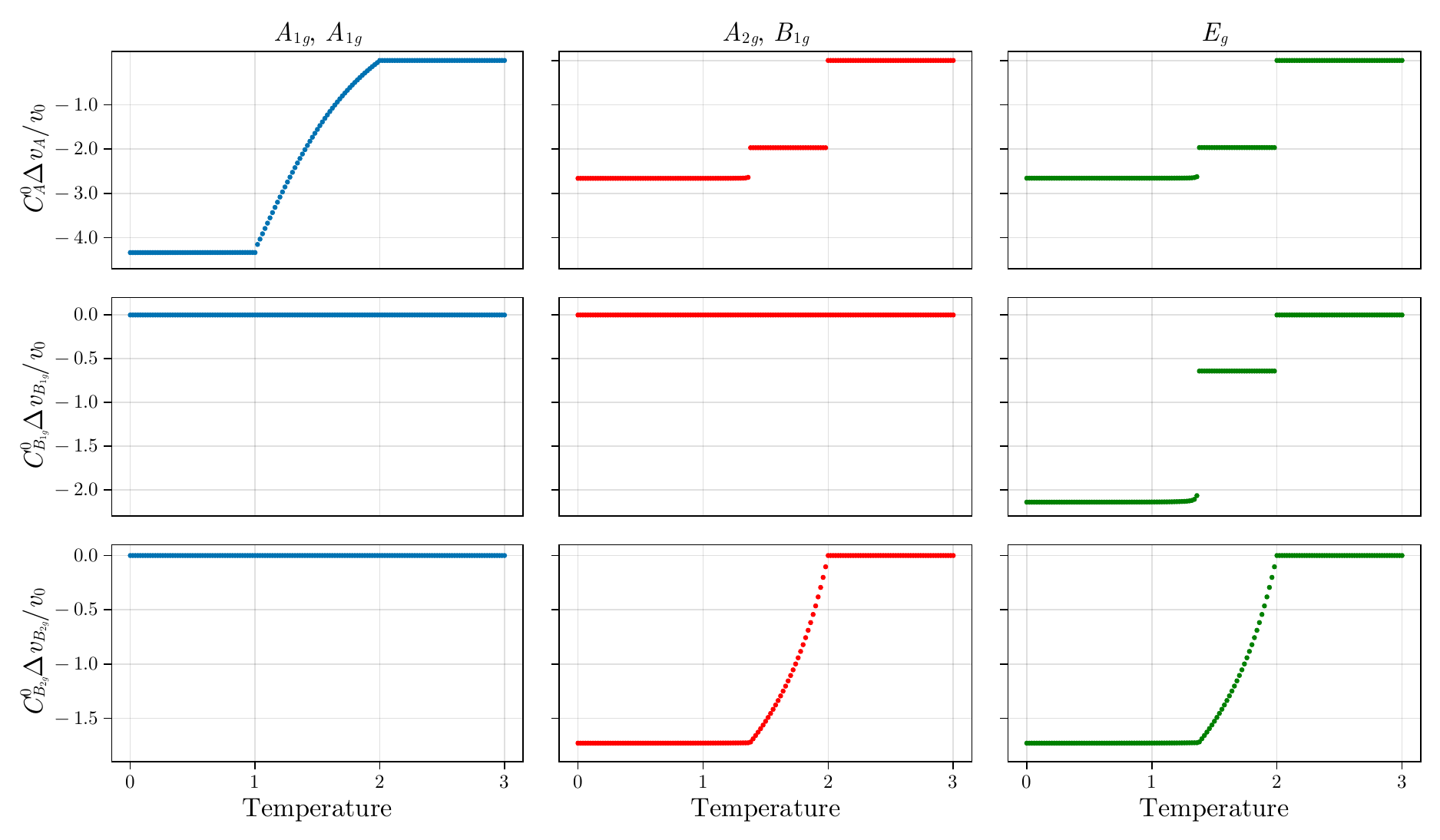}
    \caption{The change in sound velocity $v$ in different channels as a function of the temperature $T$, for the different models considered. The only difference between the models is the symmetry of the order parameter. In terms of the order parameter components, the irreps $(A_{1g},A_{1g})$ corresponds to $(s,s)$-components, $(A_{2g},B_{1g})$ to $(g,d)$-components, and $E_g$ to $(p,p)$-components. Each model gives a different ultrasound response at each phase transition. Unlike models with a single transition, the ultrasound response in the shear modes can be continuous (bottom middle, bottom right). The connection between the labels $A, B_{1g}, B_{2g}$ and the sound modes is shown in Table \ref{tab:ukuC}. }
    \label{fig:data_1}
\end{figure*}

Here, the two-component order parameter transforms as the only two-dimensional irrep of $D_{4h}$, $E_g$, often called a vector order parameter.
Examples include $p$-wave superconductors.
The symmetrized product of the order parameters transforms like $E_g \otimes E_g = A_{1g}\oplus B_{1g} \oplus B_{2g}$.
This gives rise to the order parameter bilinears $|\Psi_1|^2 + |\Psi_2|^2$, $|\Psi_1|^2 - |\Psi_2|^2$ and $\Psi_1\Psi^{\dagger}_2 + \Psi^{\dagger}_1 \Psi_2$, which belong to the $A_{1g}$, $B_{1g}$ and $B_{2g}$ irreps of $D_{4h}$, respectively.
In this case, each term pairs with a different strain, so the coupling free energy is given by
\begin{align}
    \mathcal{F}_C^{(d,d)} = \, & \gamma_1\left( |\Psi_1|^2 + |\Psi_2|^2 \right)\left( \epsilon_{xx} + \epsilon_{yy} \right) \nonumber \\
    \, & + \gamma_2\left( |\Psi_1|^2 - |\Psi_2|^2 \right) \left( \epsilon_{xx} - \epsilon_{yy} \right) \nonumber \\
    \, & + \gamma_3 \left( \Psi_1\Psi^{\dagger}_2 + \Psi^{\dagger}_1 \Psi_2 \right) \epsilon_{xy}.
\end{align}
The ultrasound response for this model is shown in the right column of Figure \ref{fig:data_1}.

This is the only model, out of the ones considered here, where there are non-zero responses in all sound modes. There are discontinuous jumps in the $A$ and $B_{1g}$ modes while the response is continuous in the $B_{2g}$ mode.

\subsection{Summary}

The ultrasound responses of the model with different order parameter symmetries are shown in Figure \ref{fig:data_1}.
Notably, the results are significantly different for the three symmetries discussed.
Hence the ultrasound is a very good tool for discriminating between possible order parameter symmetries, with different responses at both the superconducting and TRSB transitions.

The phase transitions do not always cause a discontinuous change in sound velocity.
In models with no bilinear Josephson, the response in the sound mode which couples to the bilinear Josephson term is continuous.
The presence of a bilinear Josephson term in the $A_{1g}, A_{1g}$ model softens the discontinuities in the $A$ channel.

Overall, we have found that ultrasound can be used to determine whether a theory has or doesn't have a bilinear Josephson term, and help constrain how large the coefficient is.
Unintuitively, the presence of the bilinear Josephson term has opposite effects on the ultrasound mode which couple to the term itself and the modes which couple to other quadratic terms in $\Psi_1$ and $\Psi_2$.
For instance, this can be seen in the $(A_{2g},B_{1g})$ model where the $B_{2g}$ mode is continuous when $c_1=0$, which can be seen in \eqref{eq: B2g continuous velocity change}.
However, turning on $c_1$ induces a discontinuous jump in the $B_{1g}$ mode, as observed in \eqref{eq: B1g discontinuous change}.


\section{External strain-order parameter coupling}

Direct application of the above symmetry analysis shows how external strain $\sigma^\textup{ext}$ couples to the order parameter $\boldsymbol{\Psi}$.
For now, we consider terms quadratic in the order parameter and ignore the biquadratic Josephson term, $(\Psi_1^2\Psi^{\dagger2}_2 + \Psi^{\dagger2}_1 \Psi_2^2)$.
Then there are three possible quadratic terms for a two-component model, they are given by
\begin{equation}
\label{eq:quadratics}
    |\Psi_1|^2 + |\Psi_2|^2, \quad |\Psi_1|^2 - |\Psi_2|^2, \quad \Psi_1\Psi^{\dagger}_2 + \Psi^{\dagger}_1 \Psi_2.
\end{equation}
These couple to different external strains depending on the symmetry of the order parameter.

For example, as we have seen in Sec.~\ref{subsec: A2g times B1g}, if $\Psi_1$ transforms like $B_{1g}$ and $\Psi_2$ like $A_{2g}$, there will be one quadratic term which transforms like $B_{1g}\otimes A_{2g}=B_{2g}$.
This is the bilinear Josephson term, ($\Psi_1\Psi^{\dagger}_2 + \Psi^{\dagger}_1 \Psi_2$).
The other two transform as $A_{1g}$, since $A_{2g}\otimes A_{2g} = A_{1g}$ and $B_{2g}\otimes B_{2g}=A_{1g}$.
The $A_{1g}$ terms can couple to external strain combinations which transform like $A_{1g}$, while the $B_{2g}$ term couples to the $B_{2g}$ external strain tensor.
Overall, at the level of free energy functional, these coupling terms are
\begin{align}
\label{eq: External strain-OP}
    &\left[ (|\Psi_1|^2 + |\Psi_2|^2) + (|\Psi_1|^2 - |\Psi_2|^2) \right](\sigma^\textup{ext}_{xx} + \sigma^\textup{ext}_{yy}) \nonumber  \\ 
    &+ (\Psi_1\Psi^{\dagger}_2 + \Psi^{\dagger}_1 \Psi_2)\sigma^\textup{ext}_{xy}.
\end{align}

For a $(B_{1g}, A_{2g})$ order parameter $\boldsymbol{\Psi}$, the bilinear Josephson term is not present in the potential energy $V(\boldsymbol{\Psi})$ as it transforms like $B_{2g}$.
However, most significantly, systems with no bilinear Josephson term can have one induced by an external strain.
This is because the $B_{2g}$ external strain $\sigma^\textup{ext}_{xy}$ couples to the bilinear Josephson term, giving rise to the term
\begin{equation}
    \sigma^\textup{ext}_{xy}(\Psi_1\Psi^{\dagger}_2 + \Psi^{\dagger}_1 \Psi_2)
\end{equation}
in \eqref{eq: External strain-OP}, with the external stress $\sigma^\textup{ext}_{xy}$ becoming the coefficient of the bilinear Josephson term.


\subsection{Stressed systems}

We will now consider a system under external stress.
We choose stress as it is theoretically straightforward to change the symmetry of the applied force.
A similar analysis will hold for other external fields, such as magnetic fields.
We will consider an order parameter with vector $E_g$ symmetry under no stress, $B_{1g}$ stress $\sigma^0_{B_{1g}}$ and $B_{2g}$ stress $\sigma^0_{B_{2g}}$.
We will assume that the unstressed system has equal $T_C^{U(1)}$ and $T_C^{\mathbb{Z}_2}$.

Our free energy is similar to before but with a modified potential energy,
\begin{align}
    V = \, & -a(T) \left(|\Psi_1|^2 + |\Psi_2|^2 \right) + \frac{b}{4}(|\Psi_1|^2 + |\Psi_2|^2)^2 \nonumber  \\
    \, &+ \frac{c_2}{2}\left( \Psi_1^2 \Psi^{\dagger2}_2 + \Psi^{\dagger2}_1\Psi_2^2 \right) + \sigma_{B_{1g}}^0\left( |\Psi_1|^2 -  |\Psi_2|^2 \right) \nonumber \\
    \, &+ \sigma_{B_{2g}}^0\left( \Psi_1 \Psi^{\dagger}_2 + \Psi^{\dagger}_1\Psi_2 \right).
\end{align}
Both external stresses split the phase transitions.
The $B_{1g}$ stress generates a model like the ones we've already studied, where one order parameter becomes non-zero before the other.
The $B_{2g}$ stressed system is different: both order parameters become non-zero at the same temperature.
They are first in a time-reversal symmetric phase, then switch to a broken time-reversal phase. 

\begin{figure*}[t]
    \centering
    \includegraphics[width=1.0\linewidth]{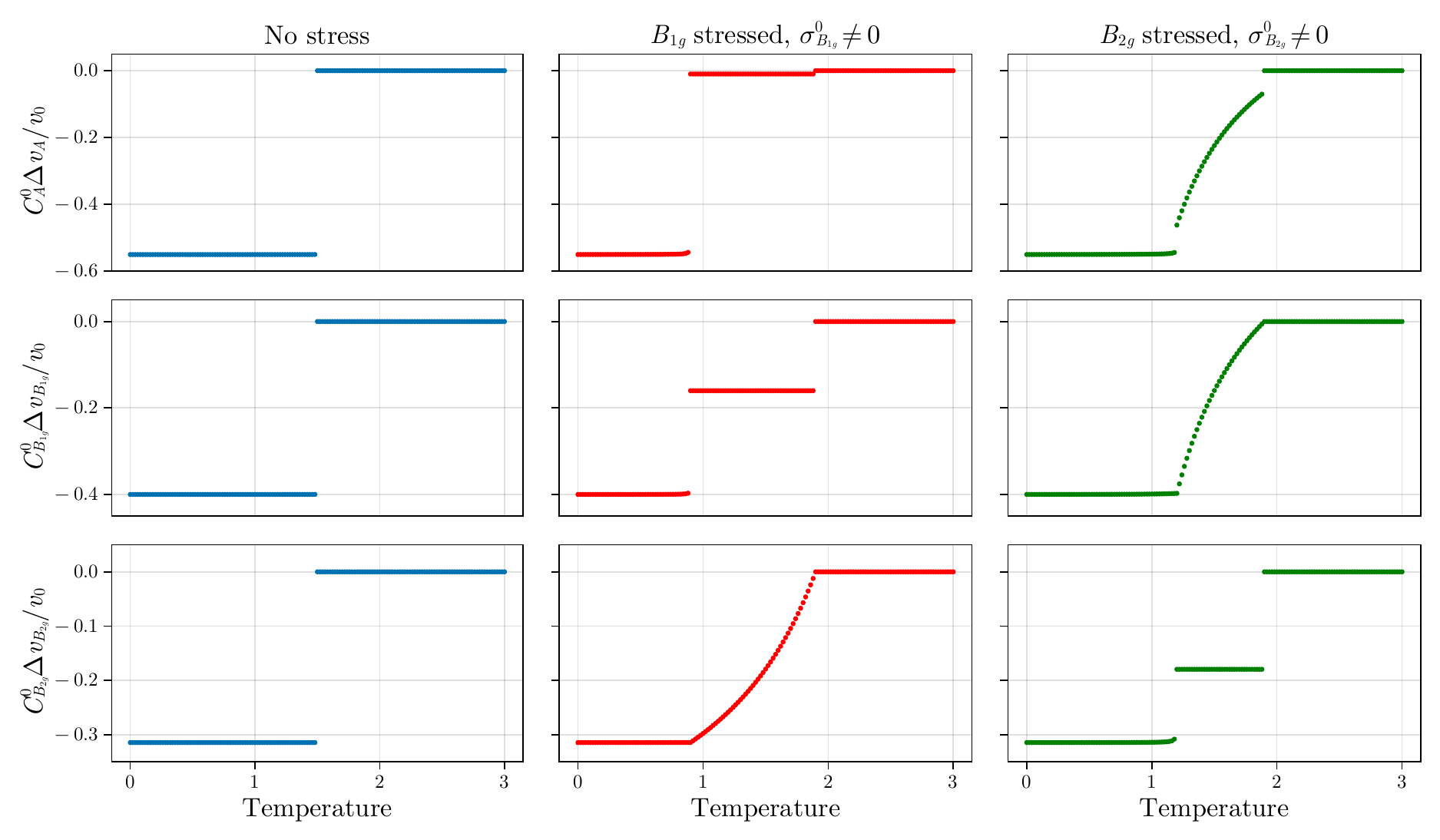}
    \caption{The ultrasound response for a non-stressed, $B_{1g}$-stressed and $B_{2g}$-stressed systems, for an $E_g$-symmetric order parameter. The ultrasound response is sensitive to the type of external stress.}
    \label{fig:data_stress}
\end{figure*}

We generate the ultrasound response for the parameters
\begin{equation*}
    \left( a(T), b, c_2, \gamma_1, \gamma_2, \gamma_3 \right) = (-T+1.5, 1, 0.4, 0.3, 0.4, 0.5)\,
\end{equation*}
in Fig. \ref{fig:data_stress}.
When they are non-zero, we choose the external stress to be $\sigma^0_a = 0.4$.
The ultrasound response is sensitive to the type of external stress applied.
Depending on the type of stress applied the jump in the ultrasound response either turns into two jumps, or becomes continuous.
When a $B_{1g}$ stress is applied, the $B_{2g}$ ultrasound response smooths out, and vice-versa.
The results show that ultrasound is sensitive to systems under different types of external stress. 

\section{Conclusion}

We have presented a theoretical framework, based on the Landau--Khalatnikov approach, to analyze ultrasound attenuation and velocity changes in superconductors that undergo TRSB transitions within the superconducting phase.
Within this framework, the order parameter dynamics near the phase transition are treated as overdamped, enabling a tractable formulation of ultrasound responce through perturbative expansions.
By incorporating the underlying crystal symmetry, strain coupling, and dynamical order parameter behavior, our model provides a detailed understanding of how different superconducting order parameter symmetries manifest in ultrasound experiments, especially focusing on the regimes where time reversal symmetry breaking occurs as a second phase transition below superconducting phase transition.
This method offers a pathway to constrain theoretical models for the order parameter symmetry of TRSB superconductors.

Our analysis shows that ultrasound responses, particularly discontinuities and smoothness in sound velocity across phase transitions, are highly sensitive to the symmetry of the superconducting order parameter and the presence of bilinear Josephson coupling.
In the considered models, the presence or absence of bilinear Josephson terms, and their symmetry, controls whether ultrasound response is continuous or discontinuous at the transitions.
In systems with intrinsic bilinear coupling, ultrasound responses can be continuous across transitions, while in systems without such coupling, responses often exhibit sharp discontinuities.
The external strain can induce a bilinear Josephson term, thereby modifying the ultrasound response and effectively tuning the symmetry of the system, which makes a combination of external static strain and sound waves an interesting probe of unconventional order parameters.

These results establish ultrasound as a precise and versatile probe for identifying both superconducting and TRSB transitions and for constraining theoretical models of unconventional superconductivity.
Our framework can be readily applied to materials of current interest, such as Sr$_2$RuO$_4$, where splitting of the BTRS and superconducting phase transition due to strain was reported in zero-field muon spin relaxation \cite{grinenko2021split}, but calorimetry has not detected two singularities \cite{jerzembeck2024t,palle2023constraints}.
Other possible material where this result may in principle be applied is Ba$_{1-x}$K$_x$Fe$_2$As$_2$. Namely
at $x\approx 0.8$ this material has a so-called quadrupling phase with time reversal symmetry breaking above the phase transition \cite{grinenko2021state,shipulin2023calorimetric}, which requires a different theory to explain ultrasound behavior, recently presented in \cite{Halcrow_2024}.
Although at adjacent doping TRSB should occur below superconducting phase transition, as shown in Fig. \ref{fig:phase_diagram}, in practice that line is extremely steep, as a function of doping $x$ in  Ba$_{1-x}$K$_x$Fe$_2$As$_2$, hence one would expect thermodynamic and ultrasound singularites easily washed out by doping inhomogeneities. 
Besides these two materials, searching second singularity below superconducting phase transition may be a useful tool for identifying new materials that break time reversal symmetry. 


\section*{Acknowledgments}
We thank Vadim Grinenko for discussions.
CH work at KTH was supported by the Carl Trygger Foundation through the grant CTS 20:25.
PL acknowledges funding from the Olle Engkvists Stiftelse through the grant 226-0103 and the Roland Gustafssons Stiftelse för teoretisk fysik.
EB was supported by the Swedish Research Council Grants  2022-04763, a project grant from Knut och Alice Wallenbergs Stiftelse,
and partially by the Wallenberg Initiative Materials Science
for Sustainability (WISE) funded by the Knut and Alice Wallenberg
Foundation.


\bibliography{main.bib}

\begin{thebibliography}{32}%
\makeatletter
\providecommand \@ifxundefined [1]{%
 \@ifx{#1\undefined}
}%
\providecommand \@ifnum [1]{%
 \ifnum #1\expandafter \@firstoftwo
 \else \expandafter \@secondoftwo
 \fi
}%
\providecommand \@ifx [1]{%
 \ifx #1\expandafter \@firstoftwo
 \else \expandafter \@secondoftwo
 \fi
}%
\providecommand \natexlab [1]{#1}%
\providecommand \enquote  [1]{``#1''}%
\providecommand \bibnamefont  [1]{#1}%
\providecommand \bibfnamefont [1]{#1}%
\providecommand \citenamefont [1]{#1}%
\providecommand \href@noop [0]{\@secondoftwo}%
\providecommand \href [0]{\begingroup \@sanitize@url \@href}%
\providecommand \@href[1]{\@@startlink{#1}\@@href}%
\providecommand \@@href[1]{\endgroup#1\@@endlink}%
\providecommand \@sanitize@url [0]{\catcode `\\12\catcode `\$12\catcode `\&12\catcode `\#12\catcode `\^12\catcode `\_12\catcode `\%12\relax}%
\providecommand \@@startlink[1]{}%
\providecommand \@@endlink[0]{}%
\providecommand \url  [0]{\begingroup\@sanitize@url \@url }%
\providecommand \@url [1]{\endgroup\@href {#1}{\urlprefix }}%
\providecommand \urlprefix  [0]{URL }%
\providecommand \Eprint [0]{\href }%
\providecommand \doibase [0]{https://doi.org/}%
\providecommand \selectlanguage [0]{\@gobble}%
\providecommand \bibinfo  [0]{\@secondoftwo}%
\providecommand \bibfield  [0]{\@secondoftwo}%
\providecommand \translation [1]{[#1]}%
\providecommand \BibitemOpen [0]{}%
\providecommand \bibitemStop [0]{}%
\providecommand \bibitemNoStop [0]{.\EOS\space}%
\providecommand \EOS [0]{\spacefactor3000\relax}%
\providecommand \BibitemShut  [1]{\csname bibitem#1\endcsname}%
\let\auto@bib@innerbib\@empty
\bibitem [{\citenamefont {Bardeen}\ \emph {et~al.}(1957)\citenamefont {Bardeen}, \citenamefont {Cooper},\ and\ \citenamefont {Schrieffer}}]{bardeen1957theory}%
  \BibitemOpen
  \bibfield  {author} {\bibinfo {author} {\bibfnamefont {J.}~\bibnamefont {Bardeen}}, \bibinfo {author} {\bibfnamefont {L.~N.}\ \bibnamefont {Cooper}},\ and\ \bibinfo {author} {\bibfnamefont {J.~R.}\ \bibnamefont {Schrieffer}},\ }\bibfield  {title} {\bibinfo {title} {Theory of superconductivity},\ }\href {https://doi.org/https://doi.org/10.1103/PhysRev.108.1175} {\bibfield  {journal} {\bibinfo  {journal} {Phys. Rev.}\ }\textbf {\bibinfo {volume} {108}},\ \bibinfo {pages} {1175} (\bibinfo {year} {1957})}\BibitemShut {NoStop}%
\bibitem [{\citenamefont {Tinkham}(2004)}]{Tinkham_2004}%
  \BibitemOpen
  \bibfield  {author} {\bibinfo {author} {\bibfnamefont {M.}~\bibnamefont {Tinkham}},\ }\href@noop {} {\emph {\bibinfo {title} {Introduction to Superconductivity}}},\ \bibinfo {edition} {2nd}\ ed.\ (\bibinfo  {publisher} {Dover Publications},\ \bibinfo {year} {2004})\BibitemShut {NoStop}%
\bibitem [{\citenamefont {Sigrist}\ and\ \citenamefont {Ueda}(1991)}]{Sigrist_1991}%
  \BibitemOpen
  \bibfield  {author} {\bibinfo {author} {\bibfnamefont {M.}~\bibnamefont {Sigrist}}\ and\ \bibinfo {author} {\bibfnamefont {K.}~\bibnamefont {Ueda}},\ }\bibfield  {title} {\bibinfo {title} {Phenomenological theory of unconventional superconductivity},\ }\href {https://doi.org/10.1103/RevModPhys.63.239} {\bibfield  {journal} {\bibinfo  {journal} {Rev. Mod. Phys.}\ }\textbf {\bibinfo {volume} {63}},\ \bibinfo {pages} {239} (\bibinfo {year} {1991})}\BibitemShut {NoStop}%
\bibitem [{\citenamefont {M{\"u}ller}\ \emph {et~al.}(1986)\citenamefont {M{\"u}ller}, \citenamefont {Maurer}, \citenamefont {Scheidt}, \citenamefont {Roth}, \citenamefont {L{\"u}ders}, \citenamefont {Bucher},\ and\ \citenamefont {B{\"o}mmel}}]{muller1986observation}%
  \BibitemOpen
  \bibfield  {author} {\bibinfo {author} {\bibfnamefont {V.}~\bibnamefont {M{\"u}ller}}, \bibinfo {author} {\bibfnamefont {D.}~\bibnamefont {Maurer}}, \bibinfo {author} {\bibfnamefont {E.-W.}\ \bibnamefont {Scheidt}}, \bibinfo {author} {\bibfnamefont {C.}~\bibnamefont {Roth}}, \bibinfo {author} {\bibfnamefont {K.}~\bibnamefont {L{\"u}ders}}, \bibinfo {author} {\bibfnamefont {E.}~\bibnamefont {Bucher}},\ and\ \bibinfo {author} {\bibfnamefont {H.}~\bibnamefont {B{\"o}mmel}},\ }\bibfield  {title} {\bibinfo {title} {Observation of a lambda-shaped ultrasonic attenuation peak in superconducting {UPt$_3$}},\ }\href {https://doi.org/https://doi.org/10.1016/0038-1098(86)90099-2} {\bibfield  {journal} {\bibinfo  {journal} {Solid State Commun.}\ }\textbf {\bibinfo {volume} {57}},\ \bibinfo {pages} {319} (\bibinfo {year} {1986})}\BibitemShut {NoStop}%
\bibitem [{\citenamefont {Golding}\ \emph {et~al.}(1985)\citenamefont {Golding}, \citenamefont {Bishop}, \citenamefont {Batlogg}, \citenamefont {Haemmerle}, \citenamefont {Fisk}, \citenamefont {Smith},\ and\ \citenamefont {Ott}}]{golding1985observation}%
  \BibitemOpen
  \bibfield  {author} {\bibinfo {author} {\bibfnamefont {B.}~\bibnamefont {Golding}}, \bibinfo {author} {\bibfnamefont {D.}~\bibnamefont {Bishop}}, \bibinfo {author} {\bibfnamefont {B.}~\bibnamefont {Batlogg}}, \bibinfo {author} {\bibfnamefont {W.}~\bibnamefont {Haemmerle}}, \bibinfo {author} {\bibfnamefont {Z.}~\bibnamefont {Fisk}}, \bibinfo {author} {\bibfnamefont {J.}~\bibnamefont {Smith}},\ and\ \bibinfo {author} {\bibfnamefont {H.}~\bibnamefont {Ott}},\ }\bibfield  {title} {\bibinfo {title} {Observation of a collective mode in superconducting {UBe$_{13}$}},\ }\href {https://doi.org/https://doi.org/10.1103/PhysRevLett.55.2479} {\bibfield  {journal} {\bibinfo  {journal} {Phys. Rev. Lett.}\ }\textbf {\bibinfo {volume} {55}},\ \bibinfo {pages} {2479} (\bibinfo {year} {1985})}\BibitemShut {NoStop}%
\bibitem [{\citenamefont {Benhabib}\ \emph {et~al.}(2021)\citenamefont {Benhabib}, \citenamefont {Lupien}, \citenamefont {Paul}, \citenamefont {Berges}, \citenamefont {Dion}, \citenamefont {Nardone}, \citenamefont {Zitouni}, \citenamefont {Mao}, \citenamefont {Maeno}, \citenamefont {Georges} \emph {et~al.}}]{benhabib2021ultrasound}%
  \BibitemOpen
  \bibfield  {author} {\bibinfo {author} {\bibfnamefont {S.}~\bibnamefont {Benhabib}}, \bibinfo {author} {\bibfnamefont {C.}~\bibnamefont {Lupien}}, \bibinfo {author} {\bibfnamefont {I.}~\bibnamefont {Paul}}, \bibinfo {author} {\bibfnamefont {L.}~\bibnamefont {Berges}}, \bibinfo {author} {\bibfnamefont {M.}~\bibnamefont {Dion}}, \bibinfo {author} {\bibfnamefont {M.}~\bibnamefont {Nardone}}, \bibinfo {author} {\bibfnamefont {A.}~\bibnamefont {Zitouni}}, \bibinfo {author} {\bibfnamefont {Z.}~\bibnamefont {Mao}}, \bibinfo {author} {\bibfnamefont {Y.}~\bibnamefont {Maeno}}, \bibinfo {author} {\bibfnamefont {A.}~\bibnamefont {Georges}}, \emph {et~al.},\ }\bibfield  {title} {\bibinfo {title} {Ultrasound evidence for a two-component superconducting order parameter in {Sr$_2$RuO$_4$}},\ }\href {https://doi.org/https://doi.org/10.1038/s41567-020-1033-3} {\bibfield  {journal} {\bibinfo  {journal} {Nat. Phys.}\ }\textbf {\bibinfo {volume} {17}},\ \bibinfo {pages} {194} (\bibinfo {year} {2021})}\BibitemShut {NoStop}%
\bibitem [{\citenamefont {Ghosh}\ \emph {et~al.}(2020)\citenamefont {Ghosh}, \citenamefont {Matty}, \citenamefont {Baumbach}, \citenamefont {Bauer}, \citenamefont {Modic}, \citenamefont {Shekhter}, \citenamefont {Mydosh}, \citenamefont {Kim},\ and\ \citenamefont {Ramshaw}}]{ghosh2020one}%
  \BibitemOpen
  \bibfield  {author} {\bibinfo {author} {\bibfnamefont {S.}~\bibnamefont {Ghosh}}, \bibinfo {author} {\bibfnamefont {M.}~\bibnamefont {Matty}}, \bibinfo {author} {\bibfnamefont {R.}~\bibnamefont {Baumbach}}, \bibinfo {author} {\bibfnamefont {E.~D.}\ \bibnamefont {Bauer}}, \bibinfo {author} {\bibfnamefont {K.~A.}\ \bibnamefont {Modic}}, \bibinfo {author} {\bibfnamefont {A.}~\bibnamefont {Shekhter}}, \bibinfo {author} {\bibfnamefont {J.}~\bibnamefont {Mydosh}}, \bibinfo {author} {\bibfnamefont {E.-A.}\ \bibnamefont {Kim}},\ and\ \bibinfo {author} {\bibfnamefont {B.}~\bibnamefont {Ramshaw}},\ }\bibfield  {title} {\bibinfo {title} {One-component order parameter in {URu$_2$Si$_2$} uncovered by resonant ultrasound spectroscopy and machine learning},\ }\href {https://doi.org/https://doi.org/10.1126/sciadv.aaz4074} {\bibfield  {journal} {\bibinfo  {journal} {Sci. Adv.}\ }\textbf {\bibinfo {volume} {6}},\ \bibinfo {pages} {eaaz4074} (\bibinfo {year} {2020})}\BibitemShut {NoStop}%
\bibitem [{\citenamefont {Theuss}\ \emph {et~al.}(2024{\natexlab{a}})\citenamefont {Theuss}, \citenamefont {Shragai}, \citenamefont {Grissonnanch}, \citenamefont {Hayes}, \citenamefont {Saha}, \citenamefont {Eo}, \citenamefont {Suarez}, \citenamefont {Shishidou}, \citenamefont {Butch}, \citenamefont {Paglione},\ and\ \citenamefont {Ramshaw}}]{Theuss_2024}%
  \BibitemOpen
  \bibfield  {author} {\bibinfo {author} {\bibfnamefont {F.}~\bibnamefont {Theuss}}, \bibinfo {author} {\bibfnamefont {A.}~\bibnamefont {Shragai}}, \bibinfo {author} {\bibfnamefont {G.}~\bibnamefont {Grissonnanch}}, \bibinfo {author} {\bibfnamefont {I.~M.}\ \bibnamefont {Hayes}}, \bibinfo {author} {\bibfnamefont {S.~R.}\ \bibnamefont {Saha}}, \bibinfo {author} {\bibfnamefont {Y.~S.}\ \bibnamefont {Eo}}, \bibinfo {author} {\bibfnamefont {A.}~\bibnamefont {Suarez}}, \bibinfo {author} {\bibfnamefont {T.}~\bibnamefont {Shishidou}}, \bibinfo {author} {\bibfnamefont {N.~P.}\ \bibnamefont {Butch}}, \bibinfo {author} {\bibfnamefont {J.}~\bibnamefont {Paglione}},\ and\ \bibinfo {author} {\bibfnamefont {B.~J.}\ \bibnamefont {Ramshaw}},\ }\bibfield  {title} {\bibinfo {title} {Single-component superconductivity in {UTe$_2$} at ambient pressure},\ }\href {https://doi.org/https://doi.org/10.1038/s41567-024-02493-1} {\bibfield  {journal} {\bibinfo  {journal} {Nat. Phys.}\ }\textbf {\bibinfo {volume} {20}},\ \bibinfo {pages}
  {1124} (\bibinfo {year} {2024}{\natexlab{a}})}\BibitemShut {NoStop}%
\bibitem [{\citenamefont {Grinenko}\ \emph {et~al.}(2021{\natexlab{a}})\citenamefont {Grinenko}, \citenamefont {Weston}, \citenamefont {Caglieris}, \citenamefont {Wuttke}, \citenamefont {Hess}, \citenamefont {Gottschall}, \citenamefont {Maccari}, \citenamefont {Gorbunov}, \citenamefont {Zherlitsyn}, \citenamefont {Wosnitza} \emph {et~al.}}]{grinenko2021state}%
  \BibitemOpen
  \bibfield  {author} {\bibinfo {author} {\bibfnamefont {V.}~\bibnamefont {Grinenko}}, \bibinfo {author} {\bibfnamefont {D.}~\bibnamefont {Weston}}, \bibinfo {author} {\bibfnamefont {F.}~\bibnamefont {Caglieris}}, \bibinfo {author} {\bibfnamefont {C.}~\bibnamefont {Wuttke}}, \bibinfo {author} {\bibfnamefont {C.}~\bibnamefont {Hess}}, \bibinfo {author} {\bibfnamefont {T.}~\bibnamefont {Gottschall}}, \bibinfo {author} {\bibfnamefont {I.}~\bibnamefont {Maccari}}, \bibinfo {author} {\bibfnamefont {D.}~\bibnamefont {Gorbunov}}, \bibinfo {author} {\bibfnamefont {S.}~\bibnamefont {Zherlitsyn}}, \bibinfo {author} {\bibfnamefont {J.}~\bibnamefont {Wosnitza}}, \emph {et~al.},\ }\bibfield  {title} {\bibinfo {title} {State with spontaneously broken time-reversal symmetry above the superconducting phase transition},\ }\href {https://doi.org/https://doi.org/10.1038/s41567-021-01350-9} {\bibfield  {journal} {\bibinfo  {journal} {Nat. Phys.}\ }\textbf {\bibinfo {volume} {17}},\ \bibinfo {pages} {1254} (\bibinfo {year}
  {2021}{\natexlab{a}})}\BibitemShut {NoStop}%
\bibitem [{\citenamefont {Halcrow}\ \emph {et~al.}(2024)\citenamefont {Halcrow}, \citenamefont {Shipulin}, \citenamefont {Caglieris}, \citenamefont {Li}, \citenamefont {Wosnitza}, \citenamefont {Klauss}, \citenamefont {Zherlitsyn}, \citenamefont {Grinenko},\ and\ \citenamefont {Babaev}}]{Halcrow_2024}%
  \BibitemOpen
  \bibfield  {author} {\bibinfo {author} {\bibfnamefont {C.}~\bibnamefont {Halcrow}}, \bibinfo {author} {\bibfnamefont {I.}~\bibnamefont {Shipulin}}, \bibinfo {author} {\bibfnamefont {F.}~\bibnamefont {Caglieris}}, \bibinfo {author} {\bibfnamefont {Y.}~\bibnamefont {Li}}, \bibinfo {author} {\bibfnamefont {J.}~\bibnamefont {Wosnitza}}, \bibinfo {author} {\bibfnamefont {H.-H.}\ \bibnamefont {Klauss}}, \bibinfo {author} {\bibfnamefont {S.}~\bibnamefont {Zherlitsyn}}, \bibinfo {author} {\bibfnamefont {V.}~\bibnamefont {Grinenko}},\ and\ \bibinfo {author} {\bibfnamefont {E.}~\bibnamefont {Babaev}},\ }\bibfield  {title} {\bibinfo {title} {Probing electron quadrupling order through ultrasound},\ }\href {https://arxiv.org/abs/2404.03020} {\bibfield  {journal} {\bibinfo  {journal} {arXiv}\ } (\bibinfo {year} {2024})},\ \Eprint {https://arxiv.org/abs/2404.03020} {arXiv:2404.03020 [cond-mat.supr-con]} \BibitemShut {NoStop}%
\bibitem [{\citenamefont {Ghosh}\ \emph {et~al.}(2022)\citenamefont {Ghosh}, \citenamefont {Kiely}, \citenamefont {Shekhter}, \citenamefont {Jerzembeck}, \citenamefont {Kikugawa}, \citenamefont {Sokolov}, \citenamefont {Mackenzie},\ and\ \citenamefont {Ramshaw}}]{ghosh2022strong}%
  \BibitemOpen
  \bibfield  {author} {\bibinfo {author} {\bibfnamefont {S.}~\bibnamefont {Ghosh}}, \bibinfo {author} {\bibfnamefont {T.~G.}\ \bibnamefont {Kiely}}, \bibinfo {author} {\bibfnamefont {A.}~\bibnamefont {Shekhter}}, \bibinfo {author} {\bibfnamefont {F.}~\bibnamefont {Jerzembeck}}, \bibinfo {author} {\bibfnamefont {N.}~\bibnamefont {Kikugawa}}, \bibinfo {author} {\bibfnamefont {D.~A.}\ \bibnamefont {Sokolov}}, \bibinfo {author} {\bibfnamefont {A.}~\bibnamefont {Mackenzie}},\ and\ \bibinfo {author} {\bibfnamefont {B.}~\bibnamefont {Ramshaw}},\ }\bibfield  {title} {\bibinfo {title} {Strong increase in ultrasound attenuation below {$T_c$} in {Sr$_2$RuO$_4$}: Possible evidence for domains},\ }\href {https://doi.org/https://doi.org/10.1103/PhysRevB.106.024520} {\bibfield  {journal} {\bibinfo  {journal} {Phys. Rev. B}\ }\textbf {\bibinfo {volume} {106}},\ \bibinfo {pages} {024520} (\bibinfo {year} {2022})}\BibitemShut {NoStop}%
\bibitem [{\citenamefont {Zhang}\ \emph {et~al.}(2019)\citenamefont {Zhang}, \citenamefont {Ding}, \citenamefont {Huang}, \citenamefont {Tan}, \citenamefont {Hillier}, \citenamefont {Biswas}, \citenamefont {MacLaughlin},\ and\ \citenamefont {Shu}}]{Zhang2019}%
  \BibitemOpen
  \bibfield  {author} {\bibinfo {author} {\bibfnamefont {J.}~\bibnamefont {Zhang}}, \bibinfo {author} {\bibfnamefont {Z.~F.}\ \bibnamefont {Ding}}, \bibinfo {author} {\bibfnamefont {K.}~\bibnamefont {Huang}}, \bibinfo {author} {\bibfnamefont {C.}~\bibnamefont {Tan}}, \bibinfo {author} {\bibfnamefont {A.~D.}\ \bibnamefont {Hillier}}, \bibinfo {author} {\bibfnamefont {P.~K.}\ \bibnamefont {Biswas}}, \bibinfo {author} {\bibfnamefont {D.~E.}\ \bibnamefont {MacLaughlin}},\ and\ \bibinfo {author} {\bibfnamefont {L.}~\bibnamefont {Shu}},\ }\bibfield  {title} {\bibinfo {title} {Broken time-reversal symmetry in superconducting {${\mathrm{Pr}}_{1\ensuremath{-}x}{\mathrm{La}}_{x}{\mathrm{Pt}}_{4}{\mathrm{Ge}}_{12}$}},\ }\href {https://doi.org/10.1103/PhysRevB.100.024508} {\bibfield  {journal} {\bibinfo  {journal} {Phys. Rev. B}\ }\textbf {\bibinfo {volume} {100}},\ \bibinfo {pages} {024508} (\bibinfo {year} {2019})}\BibitemShut {NoStop}%
\bibitem [{\citenamefont {Li}\ \emph {et~al.}(2011)\citenamefont {Li}, \citenamefont {Alidoust}, \citenamefont {Tranquada}, \citenamefont {Gu},\ and\ \citenamefont {Ong}}]{Li2011_2}%
  \BibitemOpen
  \bibfield  {author} {\bibinfo {author} {\bibfnamefont {L.}~\bibnamefont {Li}}, \bibinfo {author} {\bibfnamefont {N.}~\bibnamefont {Alidoust}}, \bibinfo {author} {\bibfnamefont {J.~M.}\ \bibnamefont {Tranquada}}, \bibinfo {author} {\bibfnamefont {G.~D.}\ \bibnamefont {Gu}},\ and\ \bibinfo {author} {\bibfnamefont {N.~P.}\ \bibnamefont {Ong}},\ }\bibfield  {title} {\bibinfo {title} {Unusual nernst effect suggesting time-reversal violation in the striped cuprate superconductor {${\mathrm{La}}_{2\ensuremath{-}x}{\mathrm{Ba}}_{x}{\mathrm{CuO}}_{4}$}},\ }\href {https://doi.org/10.1103/PhysRevLett.107.277001} {\bibfield  {journal} {\bibinfo  {journal} {Phys. Rev. Lett.}\ }\textbf {\bibinfo {volume} {107}},\ \bibinfo {pages} {277001} (\bibinfo {year} {2011})}\BibitemShut {NoStop}%
\bibitem [{\citenamefont {Xia}\ \emph {et~al.}(2006)\citenamefont {Xia}, \citenamefont {Maeno}, \citenamefont {Beyersdorf}, \citenamefont {Fejer},\ and\ \citenamefont {Kapitulnik}}]{Xia06}%
  \BibitemOpen
  \bibfield  {author} {\bibinfo {author} {\bibfnamefont {J.}~\bibnamefont {Xia}}, \bibinfo {author} {\bibfnamefont {Y.}~\bibnamefont {Maeno}}, \bibinfo {author} {\bibfnamefont {P.~T.}\ \bibnamefont {Beyersdorf}}, \bibinfo {author} {\bibfnamefont {M.~M.}\ \bibnamefont {Fejer}},\ and\ \bibinfo {author} {\bibfnamefont {A.}~\bibnamefont {Kapitulnik}},\ }\bibfield  {title} {\bibinfo {title} {High resolution polar kerr effect measurements of {${\mathrm{Sr}}_{2}{\mathrm{RuO}}_{4}$}: Evidence for broken time-reversal symmetry in the superconducting state},\ }\href {https://doi.org/10.1103/PhysRevLett.97.167002} {\bibfield  {journal} {\bibinfo  {journal} {Phys. Rev. Lett.}\ }\textbf {\bibinfo {volume} {97}},\ \bibinfo {pages} {167002} (\bibinfo {year} {2006})}\BibitemShut {NoStop}%
\bibitem [{\citenamefont {Vadimov}\ and\ \citenamefont {Silaev}(2018)}]{Vadimov2018}%
  \BibitemOpen
  \bibfield  {author} {\bibinfo {author} {\bibfnamefont {V.~L.}\ \bibnamefont {Vadimov}}\ and\ \bibinfo {author} {\bibfnamefont {M.~A.}\ \bibnamefont {Silaev}},\ }\bibfield  {title} {\bibinfo {title} {Polarization of the spontaneous magnetic field and magnetic fluctuations in $\mathit{s}+\mathit{is}$ anisotropic multiband superconductors},\ }\href {https://doi.org/10.1103/PhysRevB.98.104504} {\bibfield  {journal} {\bibinfo  {journal} {Phys. Rev. B}\ }\textbf {\bibinfo {volume} {98}},\ \bibinfo {pages} {104504} (\bibinfo {year} {2018})}\BibitemShut {NoStop}%
\bibitem [{\citenamefont {Maiti}\ and\ \citenamefont {Chubukov}(2013)}]{Maiti2013}%
  \BibitemOpen
  \bibfield  {author} {\bibinfo {author} {\bibfnamefont {S.}~\bibnamefont {Maiti}}\ and\ \bibinfo {author} {\bibfnamefont {A.~V.}\ \bibnamefont {Chubukov}},\ }\bibfield  {title} {\bibinfo {title} {$s+is$ state with broken time-reversal symmetry in {Fe}-based superconductors},\ }\href {https://doi.org/10.1103/PhysRevB.87.144511} {\bibfield  {journal} {\bibinfo  {journal} {Phys. Rev. B}\ }\textbf {\bibinfo {volume} {87}},\ \bibinfo {pages} {144511} (\bibinfo {year} {2013})}\BibitemShut {NoStop}%
\bibitem [{\citenamefont {Sigrist}(2002)}]{sigrist2002ehrenfest}%
  \BibitemOpen
  \bibfield  {author} {\bibinfo {author} {\bibfnamefont {M.}~\bibnamefont {Sigrist}},\ }\bibfield  {title} {\bibinfo {title} {Ehrenfest relations for ultrasound absorption in {Sr$_2$RuO$_4$}},\ }\href {https://doi.org/10.1143/PTP.107.917} {\bibfield  {journal} {\bibinfo  {journal} {Prog. Theor. Phys.}\ }\textbf {\bibinfo {volume} {107}},\ \bibinfo {pages} {917} (\bibinfo {year} {2002})}\BibitemShut {NoStop}%
\bibitem [{\citenamefont {Yuan}\ \emph {et~al.}(2021)\citenamefont {Yuan}, \citenamefont {Berg},\ and\ \citenamefont {Kivelson}}]{yuan2021strain}%
  \BibitemOpen
  \bibfield  {author} {\bibinfo {author} {\bibfnamefont {A.~C.}\ \bibnamefont {Yuan}}, \bibinfo {author} {\bibfnamefont {E.}~\bibnamefont {Berg}},\ and\ \bibinfo {author} {\bibfnamefont {S.~A.}\ \bibnamefont {Kivelson}},\ }\bibfield  {title} {\bibinfo {title} {Strain-induced time reversal breaking and half quantum vortices near a putative superconducting tetracritical point in {Sr$_2$RuO$_4$}},\ }\href {https://doi.org/https://doi.org/10.1103/PhysRevB.104.054518} {\bibfield  {journal} {\bibinfo  {journal} {Phys. Rev. B}\ }\textbf {\bibinfo {volume} {104}},\ \bibinfo {pages} {054518} (\bibinfo {year} {2021})}\BibitemShut {NoStop}%
\bibitem [{\citenamefont {Kaba}\ and\ \citenamefont {S\'en\'echal}(2019)}]{Kaba_2019}%
  \BibitemOpen
  \bibfield  {author} {\bibinfo {author} {\bibfnamefont {S.-O.}\ \bibnamefont {Kaba}}\ and\ \bibinfo {author} {\bibfnamefont {D.}~\bibnamefont {S\'en\'echal}},\ }\bibfield  {title} {\bibinfo {title} {Group-theoretical classification of superconducting states of strontium ruthenate},\ }\href {https://doi.org/10.1103/PhysRevB.100.214507} {\bibfield  {journal} {\bibinfo  {journal} {Phys. Rev. B}\ }\textbf {\bibinfo {volume} {100}},\ \bibinfo {pages} {214507} (\bibinfo {year} {2019})}\BibitemShut {NoStop}%
\bibitem [{\citenamefont {L\"uthi}(2005)}]{Luthi05}%
  \BibitemOpen
  \bibfield  {author} {\bibinfo {author} {\bibfnamefont {B.}~\bibnamefont {L\"uthi}},\ }\bibfield  {title} {\bibinfo {title} {Physical acoustics in the solid state},\ }\href@noop {} {\bibfield  {journal} {\bibinfo  {journal} {Springer, Heidelberg}\ } (\bibinfo {year} {2005})}\BibitemShut {NoStop}%
\bibitem [{\citenamefont {Miyake}\ and\ \citenamefont {Varma}(1986)}]{Varma_1986}%
  \BibitemOpen
  \bibfield  {author} {\bibinfo {author} {\bibfnamefont {K.}~\bibnamefont {Miyake}}\ and\ \bibinfo {author} {\bibfnamefont {C.~M.}\ \bibnamefont {Varma}},\ }\bibfield  {title} {\bibinfo {title} {{Landau-Khalatnikov} damping of ultrasound in heavy-fermion superconductors},\ }\href {https://doi.org/10.1103/PhysRevLett.57.1627} {\bibfield  {journal} {\bibinfo  {journal} {Phys. Rev. Lett.}\ }\textbf {\bibinfo {volume} {57}},\ \bibinfo {pages} {1627} (\bibinfo {year} {1986})}\BibitemShut {NoStop}%
\bibitem [{\citenamefont {Haugen}\ \emph {et~al.}(2021)\citenamefont {Haugen}, \citenamefont {Babaev}, \citenamefont {Krohg},\ and\ \citenamefont {Sudb\o{}}}]{Haugen2021first}%
  \BibitemOpen
  \bibfield  {author} {\bibinfo {author} {\bibfnamefont {H.~H.}\ \bibnamefont {Haugen}}, \bibinfo {author} {\bibfnamefont {E.}~\bibnamefont {Babaev}}, \bibinfo {author} {\bibfnamefont {F.~N.}\ \bibnamefont {Krohg}},\ and\ \bibinfo {author} {\bibfnamefont {A.}~\bibnamefont {Sudb\o{}}},\ }\bibfield  {title} {\bibinfo {title} {First-order superconducting phase transition in a chiral $p+ip$ system},\ }\href {https://doi.org/10.1103/PhysRevB.104.104515} {\bibfield  {journal} {\bibinfo  {journal} {Phys. Rev. B}\ }\textbf {\bibinfo {volume} {104}},\ \bibinfo {pages} {104515} (\bibinfo {year} {2021})}\BibitemShut {NoStop}%
\bibitem [{\citenamefont {Williams}\ and\ \citenamefont {Rudnick}(1970)}]{Williams_1970}%
  \BibitemOpen
  \bibfield  {author} {\bibinfo {author} {\bibfnamefont {R.~D.}\ \bibnamefont {Williams}}\ and\ \bibinfo {author} {\bibfnamefont {I.}~\bibnamefont {Rudnick}},\ }\bibfield  {title} {\bibinfo {title} {Attenuation of first sound near the lambda transition of liquid helium},\ }\href {https://doi.org/10.1103/PhysRevLett.25.276} {\bibfield  {journal} {\bibinfo  {journal} {Phys. Rev. Lett.}\ }\textbf {\bibinfo {volume} {25}},\ \bibinfo {pages} {276} (\bibinfo {year} {1970})}\BibitemShut {NoStop}%
\bibitem [{\citenamefont {Theuss}\ \emph {et~al.}(2024{\natexlab{b}})\citenamefont {Theuss}, \citenamefont {Simarro}, \citenamefont {Shragai}, \citenamefont {Grissonnanche}, \citenamefont {Hayes}, \citenamefont {Saha}, \citenamefont {Shishidou}, \citenamefont {Chen}, \citenamefont {Nakatsuji}, \citenamefont {Ran}, \citenamefont {Weinert}, \citenamefont {Butch}, \citenamefont {Paglione},\ and\ \citenamefont {Ramshaw}}]{Theuss_Simarro_2024}%
  \BibitemOpen
  \bibfield  {author} {\bibinfo {author} {\bibfnamefont {F.}~\bibnamefont {Theuss}}, \bibinfo {author} {\bibfnamefont {G.~d. l.~F.}\ \bibnamefont {Simarro}}, \bibinfo {author} {\bibfnamefont {A.}~\bibnamefont {Shragai}}, \bibinfo {author} {\bibfnamefont {G.}~\bibnamefont {Grissonnanche}}, \bibinfo {author} {\bibfnamefont {I.~M.}\ \bibnamefont {Hayes}}, \bibinfo {author} {\bibfnamefont {S.}~\bibnamefont {Saha}}, \bibinfo {author} {\bibfnamefont {T.}~\bibnamefont {Shishidou}}, \bibinfo {author} {\bibfnamefont {T.}~\bibnamefont {Chen}}, \bibinfo {author} {\bibfnamefont {S.}~\bibnamefont {Nakatsuji}}, \bibinfo {author} {\bibfnamefont {S.}~\bibnamefont {Ran}}, \bibinfo {author} {\bibfnamefont {M.}~\bibnamefont {Weinert}}, \bibinfo {author} {\bibfnamefont {N.~P.}\ \bibnamefont {Butch}}, \bibinfo {author} {\bibfnamefont {J.}~\bibnamefont {Paglione}},\ and\ \bibinfo {author} {\bibfnamefont {B.~J.}\ \bibnamefont {Ramshaw}},\ }\bibfield  {title} {\bibinfo {title} {Resonant ultrasound spectroscopy for irregularly shaped
  samples and its application to uranium ditelluride},\ }\href {https://doi.org/10.1103/PhysRevLett.132.066003} {\bibfield  {journal} {\bibinfo  {journal} {Phys. Rev. Lett.}\ }\textbf {\bibinfo {volume} {132}},\ \bibinfo {pages} {066003} (\bibinfo {year} {2024}{\natexlab{b}})}\BibitemShut {NoStop}%
\bibitem [{\citenamefont {Ding}\ \emph {et~al.}(2008)\citenamefont {Ding}, \citenamefont {Richard}, \citenamefont {Nakayama}, \citenamefont {Sugawara}, \citenamefont {Arakane}, \citenamefont {Sekiba}, \citenamefont {Takayama}, \citenamefont {Souma}, \citenamefont {Sato}, \citenamefont {Takahashi} \emph {et~al.}}]{ding2008observation}%
  \BibitemOpen
  \bibfield  {author} {\bibinfo {author} {\bibfnamefont {H.}~\bibnamefont {Ding}}, \bibinfo {author} {\bibfnamefont {P.}~\bibnamefont {Richard}}, \bibinfo {author} {\bibfnamefont {K.}~\bibnamefont {Nakayama}}, \bibinfo {author} {\bibfnamefont {K.}~\bibnamefont {Sugawara}}, \bibinfo {author} {\bibfnamefont {T.}~\bibnamefont {Arakane}}, \bibinfo {author} {\bibfnamefont {Y.}~\bibnamefont {Sekiba}}, \bibinfo {author} {\bibfnamefont {A.}~\bibnamefont {Takayama}}, \bibinfo {author} {\bibfnamefont {S.}~\bibnamefont {Souma}}, \bibinfo {author} {\bibfnamefont {T.}~\bibnamefont {Sato}}, \bibinfo {author} {\bibfnamefont {T.}~\bibnamefont {Takahashi}}, \emph {et~al.},\ }\bibfield  {title} {\bibinfo {title} {Observation of fermi-surface--dependent nodeless superconducting gaps in {Ba$_{0.6}$K$_{0.4}$Fe$_2$As$_2$}},\ }\href {https://doi.org/10.1209/0295-5075/83/47001} {\bibfield  {journal} {\bibinfo  {journal} {Europhys. Lett.}\ }\textbf {\bibinfo {volume} {83}},\ \bibinfo {pages} {47001} (\bibinfo {year}
  {2008})}\BibitemShut {NoStop}%
\bibitem [{\citenamefont {Nakayama}\ \emph {et~al.}(2011)\citenamefont {Nakayama}, \citenamefont {Sato}, \citenamefont {Richard}, \citenamefont {Xu}, \citenamefont {Kawahara}, \citenamefont {Umezawa}, \citenamefont {Qian}, \citenamefont {Neupane}, \citenamefont {Chen}, \citenamefont {Ding} \emph {et~al.}}]{nakayama2011universality}%
  \BibitemOpen
  \bibfield  {author} {\bibinfo {author} {\bibfnamefont {K.}~\bibnamefont {Nakayama}}, \bibinfo {author} {\bibfnamefont {T.}~\bibnamefont {Sato}}, \bibinfo {author} {\bibfnamefont {P.}~\bibnamefont {Richard}}, \bibinfo {author} {\bibfnamefont {Y.-M.}\ \bibnamefont {Xu}}, \bibinfo {author} {\bibfnamefont {T.}~\bibnamefont {Kawahara}}, \bibinfo {author} {\bibfnamefont {K.}~\bibnamefont {Umezawa}}, \bibinfo {author} {\bibfnamefont {T.}~\bibnamefont {Qian}}, \bibinfo {author} {\bibfnamefont {M.}~\bibnamefont {Neupane}}, \bibinfo {author} {\bibfnamefont {G.}~\bibnamefont {Chen}}, \bibinfo {author} {\bibfnamefont {H.}~\bibnamefont {Ding}}, \emph {et~al.},\ }\bibfield  {title} {\bibinfo {title} {Universality of superconducting gaps in overdoped {Ba$_{0.3}$K$_{0.7}$Fe$_2$As$_2$} observed by angle-resolved photoemission spectroscopy},\ }\href {https://doi.org/https://doi.org/10.1103/PhysRevB.83.020501} {\bibfield  {journal} {\bibinfo  {journal} {Phys. Rev. B}\ }\textbf {\bibinfo {volume} {83}},\ \bibinfo {pages}
  {020501} (\bibinfo {year} {2011})}\BibitemShut {NoStop}%
\bibitem [{\citenamefont {Corbae}\ \emph {et~al.}(2024)\citenamefont {Corbae}, \citenamefont {Zhang}, \citenamefont {Li}, \citenamefont {Babaev}, \citenamefont {Grinenko}, \citenamefont {Lee}, \citenamefont {Kihou}, \citenamefont {Tjernberg}, \citenamefont {Lu}, \citenamefont {Hashimoto} \emph {et~al.}}]{corbae2024fundamental}%
  \BibitemOpen
  \bibfield  {author} {\bibinfo {author} {\bibfnamefont {E.}~\bibnamefont {Corbae}}, \bibinfo {author} {\bibfnamefont {R.}~\bibnamefont {Zhang}}, \bibinfo {author} {\bibfnamefont {C.}~\bibnamefont {Li}}, \bibinfo {author} {\bibfnamefont {E.}~\bibnamefont {Babaev}}, \bibinfo {author} {\bibfnamefont {V.}~\bibnamefont {Grinenko}}, \bibinfo {author} {\bibfnamefont {C.-H.}\ \bibnamefont {Lee}}, \bibinfo {author} {\bibfnamefont {K.}~\bibnamefont {Kihou}}, \bibinfo {author} {\bibfnamefont {O.}~\bibnamefont {Tjernberg}}, \bibinfo {author} {\bibfnamefont {D.}~\bibnamefont {Lu}}, \bibinfo {author} {\bibfnamefont {M.}~\bibnamefont {Hashimoto}}, \emph {et~al.},\ }\bibfield  {title} {\bibinfo {title} {Fundamental study of hole overdoped ba1-x kx fe 2 as 2},\ }in\ \href@noop {} {\emph {\bibinfo {booktitle} {APS March Meeting Abstracts}}},\ Vol.\ \bibinfo {volume} {2024}\ (\bibinfo {year} {2024})\ pp.\ \bibinfo {pages} {M15--006}\BibitemShut {NoStop}%
\bibitem [{\citenamefont {Kivelson}\ \emph {et~al.}(2020)\citenamefont {Kivelson}, \citenamefont {Yuan}, \citenamefont {Ramshaw},\ and\ \citenamefont {Thomale}}]{kivelson2020proposal}%
  \BibitemOpen
  \bibfield  {author} {\bibinfo {author} {\bibfnamefont {S.~A.}\ \bibnamefont {Kivelson}}, \bibinfo {author} {\bibfnamefont {A.~C.}\ \bibnamefont {Yuan}}, \bibinfo {author} {\bibfnamefont {B.}~\bibnamefont {Ramshaw}},\ and\ \bibinfo {author} {\bibfnamefont {R.}~\bibnamefont {Thomale}},\ }\bibfield  {title} {\bibinfo {title} {A proposal for reconciling diverse experiments on the superconducting state in {Sr$_2$RuO$_4$}},\ }\href {https://doi.org/https://doi.org/10.1038/s41535-020-0245-1} {\bibfield  {journal} {\bibinfo  {journal} {npj Quantum Mater.}\ }\textbf {\bibinfo {volume} {5}},\ \bibinfo {pages} {43} (\bibinfo {year} {2020})}\BibitemShut {NoStop}%
\bibitem [{\citenamefont {Grinenko}\ \emph {et~al.}(2021{\natexlab{b}})\citenamefont {Grinenko}, \citenamefont {Ghosh}, \citenamefont {Sarkar}, \citenamefont {Orain}, \citenamefont {Nikitin}, \citenamefont {Elender}, \citenamefont {Das}, \citenamefont {Guguchia}, \citenamefont {Br{\"u}ckner}, \citenamefont {Barber} \emph {et~al.}}]{grinenko2021split}%
  \BibitemOpen
  \bibfield  {author} {\bibinfo {author} {\bibfnamefont {V.}~\bibnamefont {Grinenko}}, \bibinfo {author} {\bibfnamefont {S.}~\bibnamefont {Ghosh}}, \bibinfo {author} {\bibfnamefont {R.}~\bibnamefont {Sarkar}}, \bibinfo {author} {\bibfnamefont {J.-C.}\ \bibnamefont {Orain}}, \bibinfo {author} {\bibfnamefont {A.}~\bibnamefont {Nikitin}}, \bibinfo {author} {\bibfnamefont {M.}~\bibnamefont {Elender}}, \bibinfo {author} {\bibfnamefont {D.}~\bibnamefont {Das}}, \bibinfo {author} {\bibfnamefont {Z.}~\bibnamefont {Guguchia}}, \bibinfo {author} {\bibfnamefont {F.}~\bibnamefont {Br{\"u}ckner}}, \bibinfo {author} {\bibfnamefont {M.~E.}\ \bibnamefont {Barber}}, \emph {et~al.},\ }\bibfield  {title} {\bibinfo {title} {Split superconducting and time-reversal symmetry-breaking transitions in {Sr$_2$RuO$_4$} under stress},\ }\href {https://doi.org/https://doi.org/10.1038/s41567-021-01182-7} {\bibfield  {journal} {\bibinfo  {journal} {Nat. Phys.}\ }\textbf {\bibinfo {volume} {17}},\ \bibinfo {pages} {748} (\bibinfo {year}
  {2021}{\natexlab{b}})}\BibitemShut {NoStop}%
\bibitem [{\citenamefont {Jerzembeck}\ \emph {et~al.}(2024)\citenamefont {Jerzembeck}, \citenamefont {Li}, \citenamefont {Palle}, \citenamefont {Hu}, \citenamefont {Biderang}, \citenamefont {Kikugawa}, \citenamefont {Sokolov}, \citenamefont {Ghosh}, \citenamefont {Ramshaw}, \citenamefont {Scaffidi}, \citenamefont {Nicklas}, \citenamefont {Schmalian}, \citenamefont {Mackenzie},\ and\ \citenamefont {Hicks}}]{jerzembeck2024t}%
  \BibitemOpen
  \bibfield  {author} {\bibinfo {author} {\bibfnamefont {F.}~\bibnamefont {Jerzembeck}}, \bibinfo {author} {\bibfnamefont {Y.-S.}\ \bibnamefont {Li}}, \bibinfo {author} {\bibfnamefont {G.}~\bibnamefont {Palle}}, \bibinfo {author} {\bibfnamefont {Z.}~\bibnamefont {Hu}}, \bibinfo {author} {\bibfnamefont {M.}~\bibnamefont {Biderang}}, \bibinfo {author} {\bibfnamefont {N.}~\bibnamefont {Kikugawa}}, \bibinfo {author} {\bibfnamefont {D.~A.}\ \bibnamefont {Sokolov}}, \bibinfo {author} {\bibfnamefont {S.}~\bibnamefont {Ghosh}}, \bibinfo {author} {\bibfnamefont {B.~J.}\ \bibnamefont {Ramshaw}}, \bibinfo {author} {\bibfnamefont {T.}~\bibnamefont {Scaffidi}}, \bibinfo {author} {\bibfnamefont {M.}~\bibnamefont {Nicklas}}, \bibinfo {author} {\bibfnamefont {J.}~\bibnamefont {Schmalian}}, \bibinfo {author} {\bibfnamefont {A.~P.}\ \bibnamefont {Mackenzie}},\ and\ \bibinfo {author} {\bibfnamefont {C.~W.}\ \bibnamefont {Hicks}},\ }\bibfield  {title} {\bibinfo {title} {{${T}_{c}$ and the elastocaloric effect of
  ${\mathrm{Sr}}_{2}{\mathrm{RuO}}_{4}$ under $\ensuremath{\langle}110\ensuremath{\rangle}$ uniaxial stress: No indications of transition splitting}},\ }\href {https://doi.org/10.1103/PhysRevB.110.064514} {\bibfield  {journal} {\bibinfo  {journal} {Phys. Rev. B}\ }\textbf {\bibinfo {volume} {110}},\ \bibinfo {pages} {064514} (\bibinfo {year} {2024})}\BibitemShut {NoStop}%
\bibitem [{\citenamefont {Palle}\ \emph {et~al.}(2023)\citenamefont {Palle}, \citenamefont {Hicks}, \citenamefont {Valent\'{\i}}, \citenamefont {Hu}, \citenamefont {Li}, \citenamefont {Rost}, \citenamefont {Nicklas}, \citenamefont {Mackenzie},\ and\ \citenamefont {Schmalian}}]{palle2023constraints}%
  \BibitemOpen
  \bibfield  {author} {\bibinfo {author} {\bibfnamefont {G.}~\bibnamefont {Palle}}, \bibinfo {author} {\bibfnamefont {C.}~\bibnamefont {Hicks}}, \bibinfo {author} {\bibfnamefont {R.}~\bibnamefont {Valent\'{\i}}}, \bibinfo {author} {\bibfnamefont {Z.}~\bibnamefont {Hu}}, \bibinfo {author} {\bibfnamefont {Y.-S.}\ \bibnamefont {Li}}, \bibinfo {author} {\bibfnamefont {A.}~\bibnamefont {Rost}}, \bibinfo {author} {\bibfnamefont {M.}~\bibnamefont {Nicklas}}, \bibinfo {author} {\bibfnamefont {A.~P.}\ \bibnamefont {Mackenzie}},\ and\ \bibinfo {author} {\bibfnamefont {J.}~\bibnamefont {Schmalian}},\ }\bibfield  {title} {\bibinfo {title} {{Constraints on the superconducting state of ${\text{Sr}}_{2}{\text{RuO}}_{4}$ from elastocaloric measurements}},\ }\href {https://doi.org/10.1103/PhysRevB.108.094516} {\bibfield  {journal} {\bibinfo  {journal} {Phys. Rev. B}\ }\textbf {\bibinfo {volume} {108}},\ \bibinfo {pages} {094516} (\bibinfo {year} {2023})}\BibitemShut {NoStop}%
\bibitem [{\citenamefont {Shipulin}\ \emph {et~al.}(2023)\citenamefont {Shipulin}, \citenamefont {Stegani}, \citenamefont {Maccari}, \citenamefont {Kihou}, \citenamefont {Lee}, \citenamefont {Hu}, \citenamefont {Zheng}, \citenamefont {Yang}, \citenamefont {Li}, \citenamefont {Yim} \emph {et~al.}}]{shipulin2023calorimetric}%
  \BibitemOpen
  \bibfield  {author} {\bibinfo {author} {\bibfnamefont {I.}~\bibnamefont {Shipulin}}, \bibinfo {author} {\bibfnamefont {N.}~\bibnamefont {Stegani}}, \bibinfo {author} {\bibfnamefont {I.}~\bibnamefont {Maccari}}, \bibinfo {author} {\bibfnamefont {K.}~\bibnamefont {Kihou}}, \bibinfo {author} {\bibfnamefont {C.-H.}\ \bibnamefont {Lee}}, \bibinfo {author} {\bibfnamefont {Q.}~\bibnamefont {Hu}}, \bibinfo {author} {\bibfnamefont {Y.}~\bibnamefont {Zheng}}, \bibinfo {author} {\bibfnamefont {F.}~\bibnamefont {Yang}}, \bibinfo {author} {\bibfnamefont {Y.}~\bibnamefont {Li}}, \bibinfo {author} {\bibfnamefont {C.-M.}\ \bibnamefont {Yim}}, \emph {et~al.},\ }\bibfield  {title} {\bibinfo {title} {Calorimetric evidence for two phase transitions in {Ba$_{1-x}$K$_{x}$Fe$_2$As$_2$} with fermion pairing and quadrupling states},\ }\href {https://doi.org/10.1038/s41467-023-42459-0} {\bibfield  {journal} {\bibinfo  {journal} {Nat. Commun.}\ }\textbf {\bibinfo {volume} {14}},\ \bibinfo {pages} {6734} (\bibinfo {year}
  {2023})}\BibitemShut {NoStop}%
\end{thebibliography}%

\end{document}